\documentclass[traditabstract]{aa}
\usepackage{epsfig}    
\usepackage{lscape}
\usepackage{natbib}
\bibpunct{(}{)}{;}{a}{}{,}    
\usepackage{graphics}
\usepackage{graphicx,graphics}
\usepackage{color} 
\usepackage{times}
\usepackage{amsmath}
\usepackage{amssymb}
\begin{document}

\title{Two sequences of spiral galaxies with different shapes \\ of the metallicity gradients} 

\author{
         L.~S.~Pilyugin\inst{\ref{ITPA},\ref{MAO}}            \and 
         G.~Tautvai\v{s}ien\.{e}\inst{\ref{ITPA}}     
}
\institute{Institute of Theoretical Physics and Astronomy, Vilnius University, Sauletekio av. 3, 10257, Vilnius, Lithuania \label{ITPA} 
\and
  Main Astronomical Observatory, National Academy of Sciences of Ukraine, 27 Akademika Zabolotnoho St, 03680, Kiev, Ukraine \label{MAO}
  }

\abstract{
  We considered two sequences of spiral galaxies with different shapes of the radial gas-phase oxygen abundance distributions from the galaxies in the survey Mapping Nearby Galaxies
  at Apache Point Observatory (MaNGA): (1) Galaxies in which the gradient is well approximated by a single linear relation across the whole disc, that is, galaxies with an S (slope)
  gradients, (2) galaxies in which the metallicity in the inner region of the disc is at a nearly constant level  and the gradient is negative at larger radii, that is, galaxies with
  level-slope (LS) gradients. We also selected galaxies with a nearly uniform oxygen abundance  across the whole galaxy, that is, galaxies with level (L) gradients (or O/H uniform
  galaxies) with a high oxygen abundance that can be the final evolutionary stage of the two galaxy sequences described above. The radial nitrogen abundance distributions
  in galaxies with LS oxygen abundance distributions also show breaks at radii smaller than the O/H distribution breaks. The observed behaviour of the oxygen and nitrogen
  abundances with radius in these galaxies can be explained by the time delay between the nitrogen and oxygen enrichment together with the variation in the star formation history
  along the radius. These galaxies clearly show the effect of the inside-out disc evolution model, which predicts that the galactic centre evolves more rapidly than
  the regions at greater galactocentric distances. We find that the shape of the radial abundance  distribution in a galaxy is not related to its macroscopic characteristics (rotation
  velocity, stellar mass, isophotal radius, and star formation rate) and is independent of its present-day environment.
  The correlations between the gradient slopes and macroscopic characteristics of galaxies are weak in the sense that the scatter of the points in each diagram is large. 
  The galaxies with different abundance distributions (S, LS, or L) in our sample are located within the main sequence of the star-forming
  galaxies in the diagram of star formation rate -- stellar mass.  We also examined the properties of the Milky Way in the context of the considered galaxy samples. 
}


\titlerunning{Two sequences of abundance distributions in spiral galaxies}
\authorrunning{Pilyugin and Tautvai{\v s}iene}

\maketitle

\section{Introduction}

It has been known for a long time that the radial abundance gradients of the discs of spiral galaxies are negative in the sense that the abundance is higher at the centre and decreases with
galactocentric distance \citep{Searle1971,Smith1975}. The chemical enrichment of the interstellar gas in the region at a given galactocentric distance depends on the star formation
history (astration level) in the region and on the mass exchange between the region and its environment. Establishing the parameter(s) of spiral galaxies that govern the star
formation history and its variation across the discs is very important for understanding the chemical evolution of galaxies. 

A number of works have explored whether the gas-phase metallicity gradient within a galaxy is related to its stellar mass. However, contradictory results were found. \citet{Belfiore2017} 
determined the oxygen abundance distributions is a sample of star-forming galaxies of different masses using the integral field unit (IFU)
spectroscopy obtained by the survey Mapping Nearby Galaxies at Apache Point Observatory (MaNGA; \citet{Bundy2015}). They reported that the gradient is roughly flat in galaxies with
masses of 10$^{9}$~$M_{\sun}$, steepens with mass and reaches a minimum value at $\sim$3$\times$10$^{10}$~$M_{\sun}$, and then flattens as the galaxy mass increases. A similar
dependence of the metallicity gradient on stellar mass was reported by  \citet{Mingozzi2020} and \citet{Franchetto2021}.
\citet{Pilyugin2019} studied the oxygen abundance distributions for late-type MaNGA galaxies
and found that the abundance gradient is roughly constant for $M_{\star}$ $\la$ 3$\times$10$^{10}$~$M_{\sun}$, flattens near this mass, and again remains roughly constant for
higher masses.

\citet{Ho2015} considered metallicity gradients in 49 local field star-forming galaxies. They found no correlation between the metallicity gradients in dex/$R_{25}$ and stellar mass. 
\citet{Sanchez2014} examined the radial abundance gradients in a sample of galaxies measured within the  Calar Alto Legacy Integral Field Area survey (CALIFA;
\citep{Sanchez2012,Sanchez2016,GarciaBenito2015}) and found that all galaxies without clear evidence of interaction present a common gradient in the oxygen abundance,
with a characteristic slope of $-0.16$~dex/$R_{25}$  and a dispersion of  0.12~dex/$R_{25}$. \citet{SanchezMenguino2018} confirmed this result. They derived the distribution of slopes for
the oxygen abundance gradient in a sample of 95 galaxies based on the measurements at the Very Large Telescope (VLT) with the integral-field spectrograph MUSE and reported a clear peak
in the distribution that suggests a gradient that is characteristic in spiral galaxies. \citet{Pilyugin2023} found that the mean value of the gradients for a sample of 451
MaNGA galaxies and 53 nearby galaxies is $-0.20$~dex/$R_{25}$ and the scatter is 0.10~dex/$R_{25}$.

\citet{Peng2014} found that the environmental dependence of the gas metallicity can be different for star-forming centrals and star-forming satellites, that is, for all galaxy members of
groups or clusters that are not centrals. The observed difference in metallicity between star-forming centrals and star-forming satellites decreases towards high stellar masses.
\citet{Schaefer2019} investigated the environmental dependence of the gas-phase metallicity in a sample of star-forming galaxies from the MaNGA
survey. They found a small ($\sim$0.05 dex) offset in the metallicities of galaxies between satellites and centrals, but the difference in the metallicity gradients was not obvious.
\citet{Lian2019} revealed that the gas metallicity gradients of low-mass satellites flatten with increasing environmental density. \citet{Franchetto2021} found that at any given
stellar mass, the metallicity profiles of cluster galaxies are systematically flatter than those of their field counterparts. \citet{LaraLopez2022} analysed the gas metallicity gradients
for a sample of ten Fornax cluster galaxies observed with MUSE. By comparing the Fornax cluster metallicity gradients with a control sample, they found a general median offset of
$\sim$0.04~dex/$R_{e}$, and they reported that the gradients of the cluster galaxies were flatter.

\citet{Boardman2021} investigated gas-phase abundance gradients across the galaxy mass -- size plane using a sample of MaNGA galaxies. They found that the gradients varied systematically, so
that above 10$^{10}$~$M_{\sun}$, smaller galaxies displayed flatter gradients than larger galaxies at a given stellar mass. \citet{Boardman2022,Boardman2023} concluded that extended galaxies
have smooth gas-accretion histories that produce negative metallicity gradients over time, and more compact and more massive systems experienced increased merging activity
that disrupted this process, leading to flatter metallicity gradients. In other words, the authors concluded that the merging activity history determines the size of
the galaxy at a given stellar mass and its metallicity gradient, and as a consequence, the size of the galaxy at a given mass is an indicator of the metallicity gradient. 

It was found some 40 years ago  \citep[e.g.][]{VilaCostas1992,Zaritsky1992} that there are breaks in the oxygen abundance gradients in some spiral galaxies.
\citet{SanchezMenguino2016,SanchezMenguino2018} found that in addition to the general negative gradient, an inner drop and an outer flattening can be observed in the radial profile of the oxygen
abundance. \citet{Belfiore2017,Belfiore2019} noted that the shape of the gradient can be a function of the stellar mass of the galaxy in the sense that the massive
galaxies ($M_{\star}$ $\ga$ 3$\times$10$^{10}$~$M_{\sun}$) are characterised by a flattening of the metallicity gradient in the central regions. 

It was suggested that the change in the slopes of the oxygen abundance gradients in spiral galaxies can be attributed to the impact of galactic barred structures on the abundance
distribution in the interstellar medium \citep[e.g.][]{VilaCostas1992,Friedli1994,Martel2013,Zurita2021}.
However, the influence of the bar (if exists) on the abundance distribution is not evident. \citet{Sanchez2016} found that the slope of the abundance gradient is independent of the presence of
the bar. 
\citet{Chen2023} investigated the metallicity gradients across the disc of five nearby barred galaxies. They observed cases with a shallow-steep metallicity radial profile, with evidence
that the bar flattens the metallicity gradients in the inner region of the galaxy, and they
also found patterns in which the bar appears to drive a steeper metallicity gradient in the central region of galaxy, producing steep-shallow metallicity profiles. 

The reported correlations between metallicity gradient and stellar mass (and other macroscopic characteristics) are fuzzy in the sense that the scatter in the diagram of
gradient versus stellar mass is large. Some fraction of the scatter in the metallicity gradients can be attributed to the
uncertainties in the gradient determinations.  Here we select and examine the well-measured spiral galaxies from the MaNGA survey.
The selection (selection criteria) of the galaxies that we analysed is described in Section 2.1. 
The examination of these well-measured galaxies holds information on the real scatter in the diagram of the gradient -- stellar mass (and other macroscopic
parameters).

We selected galaxies for which the radial oxygen abundance distribution is either perfectly approximated by the straight line across the whole disc or for which the metallicity
in the inner region of the disc is at a nearly constant level and the gradient is negative at larger radii. These galaxies can be considered representatives of two alternative
sequences of the abundance distributions in galaxies.
We also selected galaxies throughout which the oxygen abundance is nearly uniform, which are the galaxies with the flattest gradients. They can be a limiting case of the two galaxy sequences
described above. We investigated the relations between the shape of the radial abundance distribution in a galaxy and its macroscopic characteristics (rotation velocity,
stellar mass, isophotal radius, and star formation rate). The Milky Way is one of the rare galaxies with very steep gradients. It is of particular interest to compare the Milky Way with
the galaxies that have flat gradients that we selected here. 

The paper is organised in the following way: The data are described in Section 2, Section 3 includes the discussion, and Section 4 contains a brief summary.

We specify the position of the point in the disc by the fractional radius $R_{g}$ normalised to the isophotal (or optical) radius  $R_{g}$ = $R/R_{25}$,  
 which is the radial distance of the isophote at a surface brightness of 25 mag$_{B}$ arcsec$^{-2}$, corrected for the galaxy inclination. The determination of
the optical radii for our sample of galaxies is described in Section 2.2.1.  


\section{Two sequences of spiral galaxies}

\subsection{Sample selection}

Our investigation is based on galaxies from the MaNGA survey (\citealt{Bundy2015}). 
Based on the publicly available spectroscopy obtained by the MaNGA survey (Data Release 13, DR13,  \citealt{Albareti2017} and  DR15 \citealt{Aguado2019}), the properties of
several hundred MaNGA galaxies were examined using our own line measurements \citep{Pilyugin2018,Pilyugin2019,Pilyugin2020,Pilyugin2021}.
For the current study, we selected a sample of galaxies by visual inspection using the criteria described below.

We selected galaxies for which the curves of the isovelocities in the measured gas velocity fields resemble a set of parabola-like curves (hourglass-like appearance of the rotation disc).
This condition provides the possibility to determine the orientation of the galaxy in space (the pixel coordinates of the rotation centre of a galaxy, the position angle of
the major kinematic axis, and the inclination angle of a galaxy) and the rotation curve. This criterion also rejects strongly interacting and merging galaxies. 

Galaxies with an inclination angle larger than $\sim$70$\degr$ were rejected. On the one hand, a fit of the H$\alpha$ velocity field in galaxies with a high ratio of
the major to minor axis can produce unrealistic values of the inclination angle \citep{Epinat2008}, and consequently, the esimated galactocentric distances of the spaxels located
far from the major axis can involve large uncertainties. On the other hand, the interpretation of the abundance in an individual spaxel is not beyond question because the spaxel
spectra involve radiations that originated at different galactocentric distances. 

We considered MaNGA galaxies mapped with 91 and 127 fibre IFU, covering  27$\farcs$5 and 32$\farcs$5 on the sky (with a large number of spaxels over the galaxy image).
We only considered galaxies for which the spaxels with measured emission lines are well distributed across galactic discs and cover more than $\sim 0.8~R_{25}$. This condition
provides the possibility to estimate reliable values of the kinematic angles and the rotation curve, the optical radius, and the shapes of the radial distributions of the oxygen
and nitrogen abundances. Our total sample involves 136 galaxies. 

The completed observations of MaNGA galaxies were included in Data Release 17 \citep{Abdurrouf2022}. The MaNGA data products were also revised for all the observations previously
released in DR15 and before (e.g. the flux calibration was updated). The emission line parameters of the spaxel spectra for our sample of galaxies are available from the MaNGA Data Analysis
Pipeline (DAP) measurements. To have uniform data, we re-derived the characteristics of our sample of 136 galaxies using the last version of DAP measurements
manga-n-n-MAPS-SPX-MILESHC-MASTARSSP.fits.gz\footnote{https://data.sdss.org/sas/dr17/manga/spectro/analysis/v3\_1\_1/3.1.0/SPX-MILESHC-MASTARSSP/} for the spectral measurements and
the Data Reduction Pipeline (DRP) measurements manga-n-n-LOGCUBE.fits.gz\footnote{https://dr17.sdss.org/sas/dr17/manga/spectro/redux/v3\_1\_1/} for the photometric data.
We discuss the determinations of the geometrical galaxy parameters (the coordinates  of the centre, the position angle of the major axis, the inclination angle, and the isophotal radius) and
the radial distributions of characteristics of the galaxies in the next section.

\begin{figure*}
\resizebox{1.00\hsize}{!}{\includegraphics[angle=000]{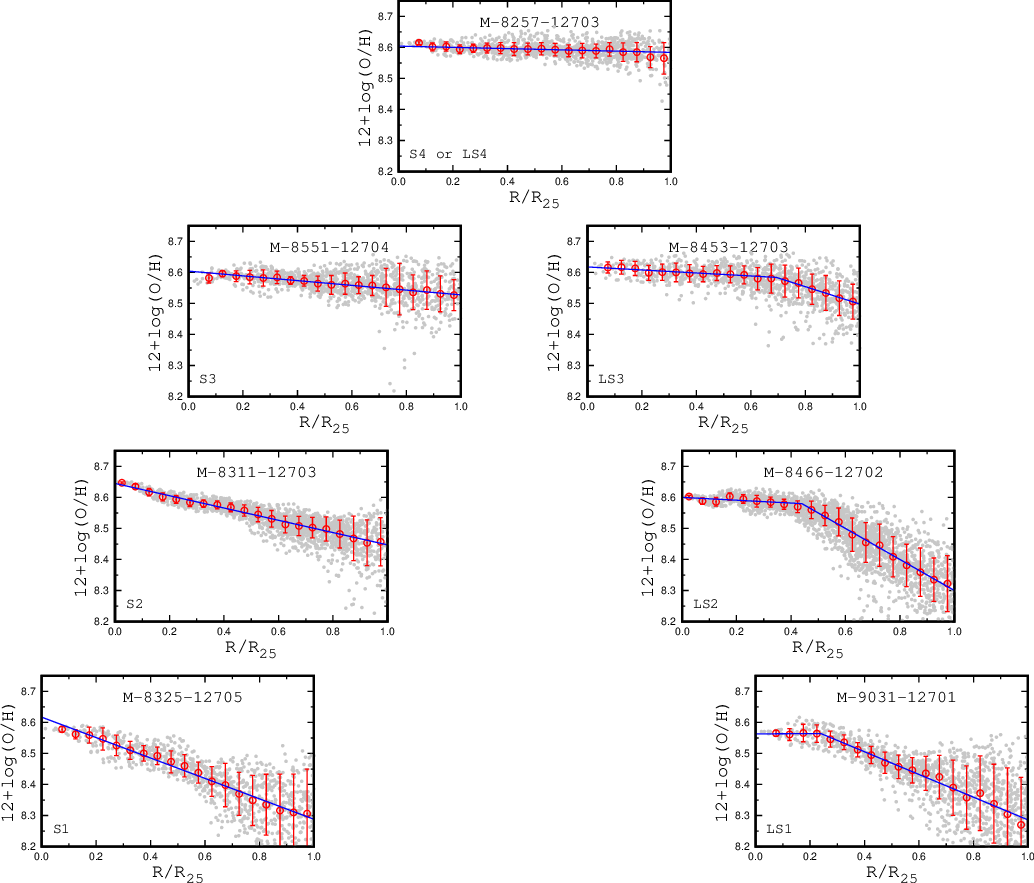}}
\caption{
  Two sequences of galaxies with different shapes of the metallicity gradients. Each panel shows the radial oxygen abundance distribution in the galaxy. The left panels show
  the galaxies for which the (negative)  gradient is well approximated by a
  single linear relation across the whole disc, i.e. the S (slope) gradient. The right panels show the galaxies in which the metallicity in the inner region of the disc ($R<R_{b,{\rm OH}}$) is at
  (nearly) constant level and the gradient is negative at larger radii, the LS gradient.
  The grey points denote the abundances for the individual spaxels, the red circles are the median values of
  the O/H in bins of 0.05 in the fractional radius $R/R_{25}$, and the bars show the scatter in O/H about the median value in the bins. The line shows the adopted relation for the radial
  abundance distribution.
  The gradients flatten along the sequence of galaxies with the S gradients:  $-0.329$~dex/$R_{25}$ in S1 (M-8325-12705),  $-0.199$~dex/$R_{25}$ in S2 (M-8311-12703), and  $-0.075$~dex/$R_{25}$
  in S3 (M-8551-12704). 
  The size of the inner region with a constant level of abundance $R_{b,OH}$ increases along the sequence of galaxies with the LS gradients:  0.24~$R_{25}$ in LS1 (M-9031-12701),  0.44~$R_{25}$
  in  LS2 (M-8466-12702), and 0.69~$R_{25}$ in LS3 (M-8453-12703).
  Galaxy M-8257-12703 can be considered either as the  limiting case of the sequence of galaxies with the S gradients (gradient is close to zero, grad =  $-0.019$~dex/$R_{25}$) or as
  the  limiting case of the sequence of galaxies with an LS gradients ($R_{b,{\rm OH}}$ =  1.0~$R_{25}$).  
}
\label{figure:schema}
\end{figure*}

We examined the re-derived radial oxygen abundance distributions in our total sample of 136 galaxies. It is known that the radial abundance distribution in many galaxies is approximated
satisfactorily by a single linear relation, and the scatter around the linear relation is usually within 0.05~dex \citep{Zinchenko2016,Pilyugin2017}. We selected a sample of galaxies
in which the (negative) gradient is approximated by a single linear relation across the whole disc with a mean value of the deviation from the relation lower than $\sim$0.01~dex
for the binned oxygen abundance (see Section 2.2.2).
This strong restriction on the scatter in abundances around the O/H -- R relation, coupled with the above selection criterion that the spaxels with measured abundances should be well
distributed across the galactic disc, allows us to hope that we selected galaxies with well-determined metallicity gradients. These galaxies are referred to as galaxies with a slope-type (S-type)
radial abundance distribution. The sequence of galaxies of the S-type abundance distribution with different values of the slope is shown in the left panels of  Fig.~\ref{figure:schema}. 
 
We also selected another sample of galaxies in which the metallicity in the inner region of the disc ($R$ $<$ $R_{b,{\rm OH}}$) is at a (nearly) constant level and the gradient is negative at
larger radii. We assumed that the oxygen abundance within the region is at a (nearly) constant level when the gradient is flatter than $\sim -0.05$~dex/$R_{25}$. This limiting value of
the slope for the selection of the regions with a uniform (constant) oxygen abundance corresponds to the uncertainty in the radial oxygen abundance gradient determined from MaNGA
measurements, which was estimated through a comparison of the values of the oxygen abundance gradient determined using three independent
MaNGA observations of the same galaxy (Section 2.2.4). The selected galaxies are referred to as  galaxies with a level-slope (LS) radial abundance distribution.
The galaxy sequence of the LS-type radial abundance distribution with increasing $R_{b,{\rm OH}}$ values is shown in the right panels of  Fig.~\ref{figure:schema}.

The galaxies in which the oxygen abundance is at a (nearly) constant level across the whole disc are referred to as galaxies with a level-type (L-type) radial abundance distribution.
These galaxies can be considered as the limiting case of the two galaxy sequences with an S-type abundance distribution (slope close to zero)
and of the galaxy sequence with an LS-type abundance distribution (the size of the region with constant level of abundance is close to the isophotal radius of the
galaxy,  $R_{b,{\rm OH}}$ $\sim$ 1.0~$R_{25}$); see the schema in Fig.~\ref{figure:schema}.

Thus, we selected 29 galaxies with an S-type abundance distribution, 23 galaxies with an LS-type abundance distribution, and 8 galaxies with an L-type abundance distribution. We investigated
the relations between the abundances and macroscopic properties of the galaxies with these abundance distributions.

\subsection{Determinations of the radial distributions of the characteristics across the galaxies}

\subsubsection{Rotation curve and  photometric profile}

\begin{figure}
\resizebox{1.00\hsize}{!}{\includegraphics[angle=000]{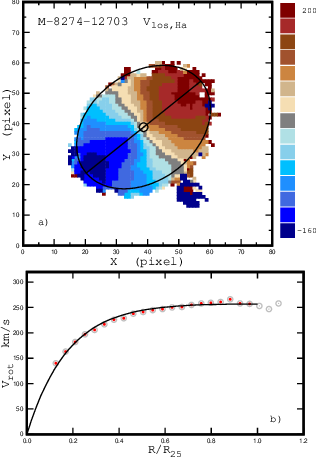}}
\caption{
Rotation of the MaNGA galaxy M-8274-12703. 
{\em Panel} {\bf a:} Line-of-sight velocity field in sky coordinates (pixel). The velocity value is colour-coded. The circle shows the kinematic centre of the galaxy,
the line indicates the position of the major kinematic axis of the galaxy, and the ellipse is its optical radius.
{\em Panel} {\bf b:} Rotation curve (points) and its fit (line) within the optical radius. 
}
\label{figure:m-8274-12703-rc}
\end{figure}

\begin{figure}
\resizebox{1.00\hsize}{!}{\includegraphics[angle=000]{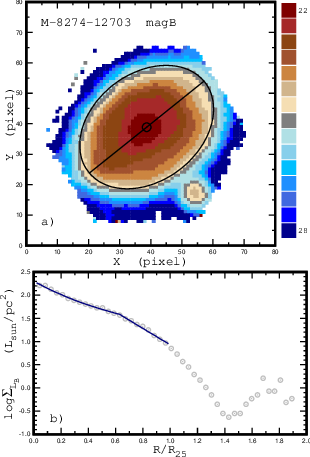}}
\caption{
Map of the surface brightness and photometric profile of the MaNGA galaxy M-8274-12703.
{\em Panel} {\bf a:} Observed surface brightness distribution across the image of the galaxy in sky coordinates (pixels). The surface brightness value is colour-coded.
The circle shows the kinematic centre of the galaxy, the line indicates the position of the major kinematic axis of the galaxy, and the  ellipse is its 
optical radius. {\em Panel} {\bf b:} Photometric profile (points) and its fit (line) within the optical radius. The photometric profile is obtained for the kinematic angles
and corrected for the galaxy inclination. 
}
\label{figure:m-8274-12703-photometry}
\end{figure}

\begin{figure}
\resizebox{1.00\hsize}{!}{\includegraphics[angle=000]{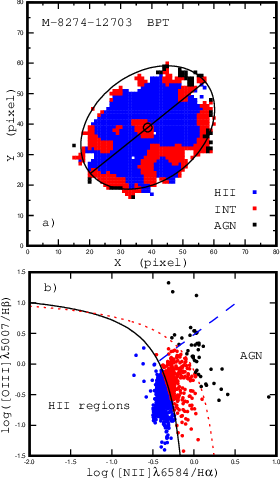}}
\caption{
BPT types of the spaxel spectra of the MaNGA galaxy M-8274-12703.
{\em Panel} {\bf a:} Map of the BPT types of the spaxel spectra. The BPT radiation types for individual spaxels are colour-coded. The circle shows the kinematic
centre of the galaxy, the line indicates the position of the major kinematic axis of the galaxy, and the ellipse is the optical radius. 
{\em Panel} {\bf b:}  BPT diagram for the individual spaxels with H\,{\sc ii}-region-like (blue), intermediate (red), and AGN-like (black) spectrum classification.
The solid and short-dashed curves mark the demarcation line between AGNs and H\,{\sc ii} regions defined by \citet{Kauffmann2003}
and \citet{Kewley2001}, respectively. The long-dashed line is the dividing line between Seyfert galaxies and LINERs defined by \citet{CidFernandes2010}.
}
\label{figure:m-8274-12703-bpt}
\end{figure}

\begin{figure}
\resizebox{1.00\hsize}{!}{\includegraphics[angle=000]{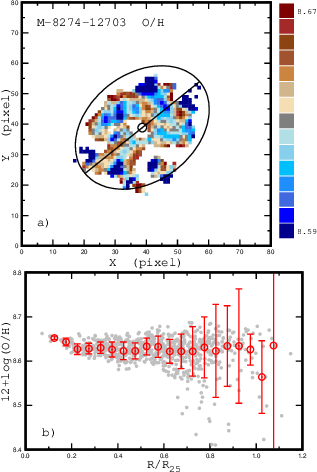}}
\caption{
Oxygen abundance in the MaNGA galaxy M-8274-12703.
{\em Panel} {\bf a:} Oxygen abundance map. The distribution across the image of the galaxy in sky coordinates (pixels). The oxygen abundance value is colour-coded.
The circle shows the kinematic centre of the galaxy, the line indicates the position of the major kinematic axis of the galaxy, and the ellipse is the 
optical radius. 
{\em Panel} {\bf b:} Radial oxygen abundance distribution. The grey points denote the abundances for the individual spaxels, the red circles are the median values of the O/H in bins
of 0.05 in the fractional radius $R/R_{25}$, and the bars show the scatter in the O/H about the median value in the bins. 
}
\label{figure:m-8274-12703-oh-map}
\end{figure}

\begin{figure}
\resizebox{1.00\hsize}{!}{\includegraphics[angle=000]{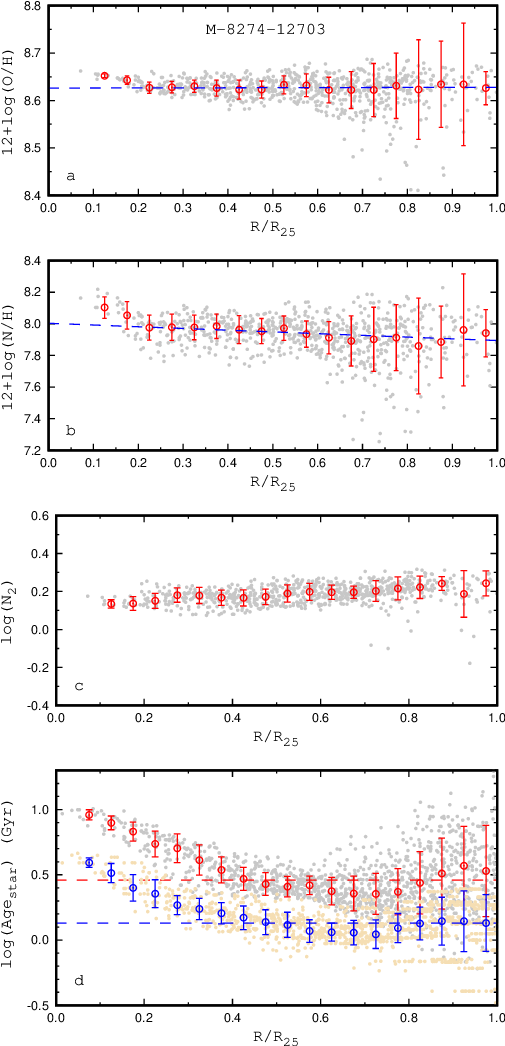}}
\caption{
  M-8274-12703 is an example of a galaxy with an L-type abundance distribution. 
  {\em Panel} {\bf a:} Radial oxygen abundance distribution. The grey points denote the abundances for the individual spaxels, and the red circles are the median values of the O/H in bins
  of 0.05~dex in the fractional radius $R/R_{25}$. The line shows the relation we adopted for the radial abundance distribution.  
  {\em Panel} {\bf b:} Same as panel (a), but for the nitrogen abundance. 
  {\em Panel} {\bf c:} Intensity of the emission nitrogen line N$_{2}$ as a function of radius. The notations are similar to those in panel a. 
  {\em Panel} {\bf d:} Age of the stellar population as a function of radius. The grey points denote the ages of the stellar populations in the individual spaxels estimated from the D$_{n}$(4000) 
  index, the red circles mark the median values of the ages in bins, and the dashed red line shows
  the median value of the ages in all the spaxels within optical radius.
  The yellow points denote the luminosity-weighed ages of the stellar populations in the individual spaxels taken from \citet{Sanchez2022},  
  the blue circles mark the median values of the ages in bins, and the dashed blue line shows
  the median value of the ages in all the spaxels within the optical radius.
}
\label{figure:m-8274-12703-r-x}
\end{figure}

\begin{figure}
\resizebox{1.00\hsize}{!}{\includegraphics[angle=000]{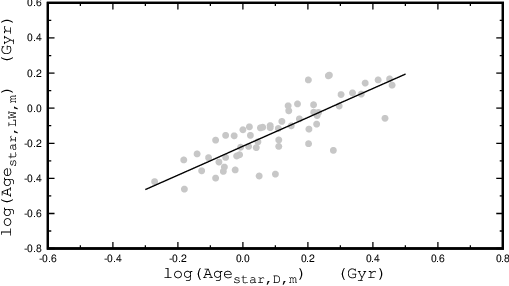}}
\caption{
  Comparison of the ages of the stellar population in a galaxy estimated in two ways.
  The median value of the luminosity-weighed ages in all the spaxels within the optical radius vs median value of ages estimated from the D$_{n}$(4000) index.
  The points denote individual MaNGA galaxy of our sample, and the line is the best fit to these points.
}
\label{figure:aged-agelw}
\end{figure}

\begin{figure}
\resizebox{1.00\hsize}{!}{\includegraphics[angle=000]{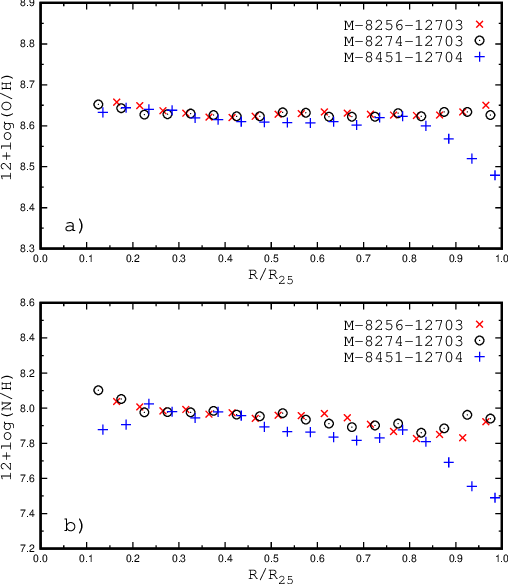}}
\caption{
  Comparison of the radial distributions of the oxygen  ({\em panel} {\bf a}) and nitrogen ({\em panel} {\bf b}) abundances in a galaxy determined from three
  independent MaNGA observations of the same galaxy.
}
\label{figure:m-8274-12703-oh-nh}
\end{figure}

\begin{figure}
\resizebox{1.00\hsize}{!}{\includegraphics[angle=000]{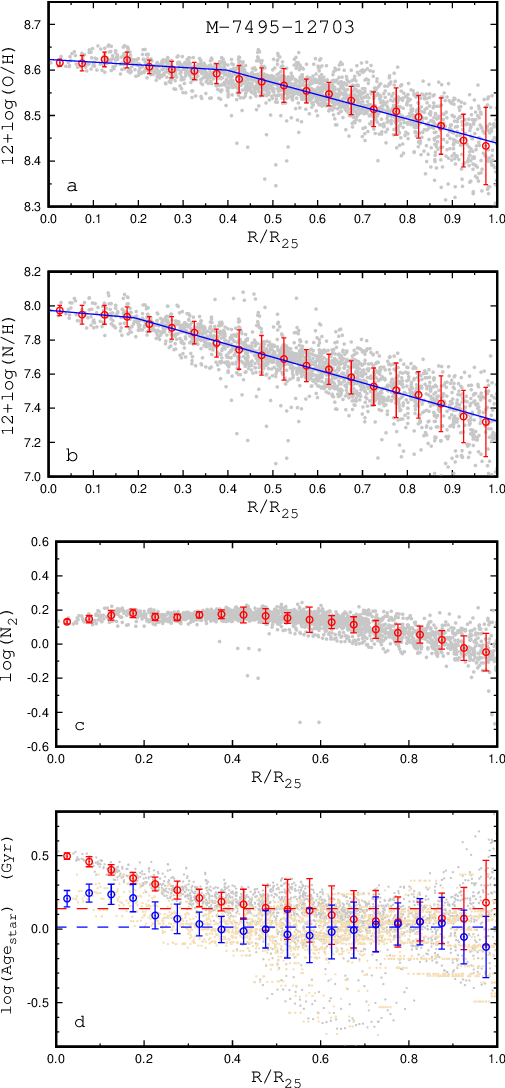}}
\caption{
  M-7495-12703 is an example of a galaxy with an LS-type abundance distribution. The notations are the same as in Fig.~\ref{figure:m-8274-12703-r-x} 
}
\label{figure:m-7495-12703-r-x}
\end{figure}

\begin{figure}
\resizebox{1.00\hsize}{!}{\includegraphics[angle=000]{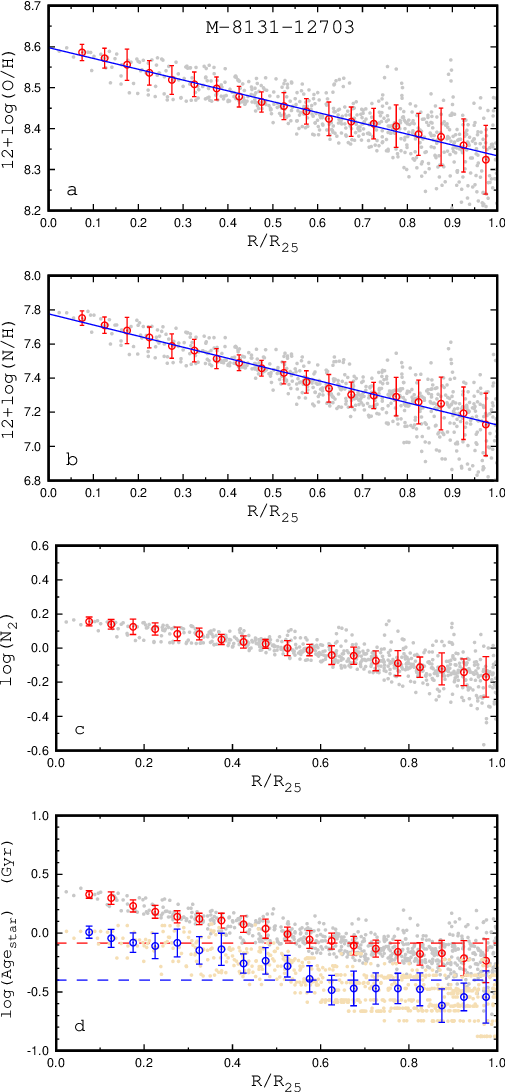}}
\caption{
  M-8131-12703 is an example of a galaxy with an S-type abundance distribution. The notations are the same as in Fig.~\ref{figure:m-8274-12703-r-x} 
}
\label{figure:m-8131-12703-r-x}
\end{figure}

\begin{figure}
\resizebox{1.00\hsize}{!}{\includegraphics[angle=000]{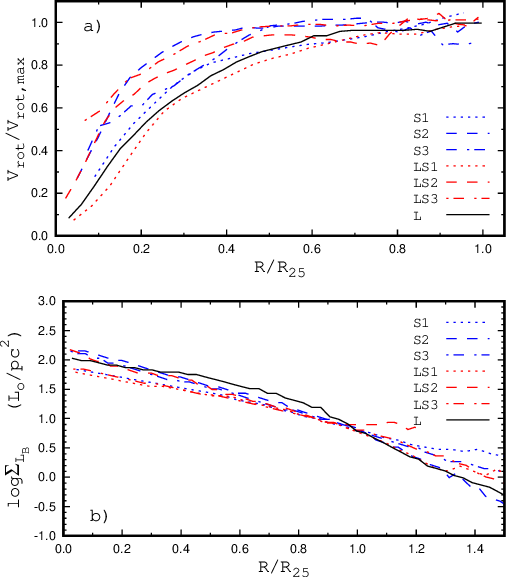}}
\caption{
  Rotation curves normalised to the maximum value of the rotation velocity ({\em panel} {\bf a}) and the photometric profiles ({\em panel} {\bf b})
  for the galaxies from  Fig.~\ref{figure:schema}.  
}
\label{figure:rc-sb-schema}
\end{figure}

The geometrical parameters of the galaxy (the coordinates  of the centre, the position angle of the major axis, the inclination angle, and the isophotal radius) are necessary
to determine the galactocentric distances of individual spaxels and to construct the radial distributions of the characteristics. 
The pixel coordinates ($x_{0}$ and $y_{0}$) of the rotation centre of the galaxy, the position angle $PA$ of the major axis, the inclination angle $i$, and the rotation
curve were derived through the best fit to the observed line-of-sight velocity field (obtained from the measured wavelength of the emission  H$\alpha$ line in the spaxel spectra)
using an iterative procedure. The detailed description is reported in \citet{Pilyugin2019}. 
In brief, in the first step, the values of $x_{0}$, $y_{0}$, $PA$, $i$, and the rotation curve are determined using all the available line-of-sight velocity measurements.
At each step, data points with large deviations from the rotation curve obtained in the previous step are rejected, and new values of the parameters and
of the rotation curve  $V_{rot}(R)$ are derived. The iteration is stopped when the absolute values of the difference of $x_{0}$ (and $y_{0}$) obtained in successive steps are lower than
0.1 pixels, the difference of $PA$ (and $i$) is smaller than 0$\fdg$1, and the rotation curves agree within 1~km\,s$^{-1}$ (at each radius).  The iteration converges after about ten steps.

The line-of-sight velocity field in sky coordinates and the derived rotation curve (points) for the MaNGA galaxy M-8274-12703 are shown
  in panels (a) and (b) of Fig.~\ref{figure:m-8274-12703-rc}, respectively.
The galactocentric distances of the spaxels determined with the coordinates of the rotation centre of the galaxy and
angles $PA$ and $i$ are used below to construct the radial distributions of the oxygen and nitrogen abundances and other characteristics across the disc.

We specified the rotation of the galaxy by the maximum value of the rotation velocity within the optical radius. To estimate this value, we carried out a least-squares fit of the
rotation curve within the optical radius using an empirical Polyex curve \citep{Giovanelli2002,Spekkens2005},
\begin{equation}
V_{pe}(r)   =  V_{o} (1 - e^{-r/r_{pe}})(1+\beta \, r/r_{pe})   ,
\label{equation:vpe}
\end{equation}
where $V_{0}$ sets the amplitude of the fit, $r_{pe}$ is the scale length that governs the inner rotation curve slope, and $\beta$ determines the slope of the outer rotation curve.
The solid line in panel (b) of Fig.~\ref{figure:m-8274-12703-rc} is a fit with the Polyex curve. The Polyex curve is a good approximation for the rotation curves with simple shapes,
but is not a satisfactory approximation of complex rotation curves, such as rotation curves with a hump. In this case, the rotation curve fit results in an incorrect maximum value of the
rotation velocity  within the optical radius. To obtain the correct maximum value of the rotation velocity within the optical radius in these cases, we did not fit the whole rotation curve
with a Polyex curve, but only the outer part of the rotation curve (circles marked by the red points in panel (b) of Fig.~\ref{figure:m-8274-12703-rc}).

  We specified the galaxy size by the optical (or isophotal) radius $R_{25}$, which is the radial distance of the isophote at the surface brightnes of 25 mag$_{B}$ arcsec$^{-2}$, corrected
  for the galaxy inclination. The measurements in the SDSS filters $g$ and $r$ for each spaxel were converted into $B$-band magnitudes following \citet{Pilyugin2018}. The observed surface
  brightness distribution across the image of the MaNGA galaxy M-8274-12703 in sky coordinates (pixels) is shown in panel (a) of Fig.~\ref{figure:m-8274-12703-photometry}. 
  The radial surface brightness distribution was constructed using the galactocentric distances of the spaxels determined with the coordinates of the centre, the position angle of the
  major axis, and the inclination angle obtained from the analysis of the observed velocity field. The surface brightness distribution was fitted by a broken exponential profile
  for the disc and by a general Sérsic profile for the bulge. The value of the isophotal radius $R_{25}$ was estimated using the fit corrected for the galaxy inclination.
  We used a simple correction for the inclination (factor cos$i$) for the surface brightness of the disc component. The bulge was assumed to be spherical, and its surface brightness was not
  corrected for inclination. Because we are interested in the surface brightness of the outer galaxy regions (determination of the optical radius), this approach is justified
  even when the bulge is not spherical. The photometric profile (points) and its fit (line) within the optical radius for the MaNGA galaxy M-8274-12703 are shown in panel (b) of
  Fig.~\ref{figure:m-8274-12703-photometry}.

\subsubsection{Radial distributions of the oxygen and nitrogen abundances}

The measured line fluxes in the spaxel spectra were corrected for the interstellar reddening using the reddening law of \citet{Cardelli1989} with $R_{V}$ = 3.1. 
The logarithmic extinction at H$\beta$ was estimated through a comparison of the measured and theoretical $F_{{\rm}H\alpha}/F_{{\rm H}\beta}$ ratios, where the theoretical
value of the line ratio (= 2.87) was taken from \citet{Osterbrock2006}, assuming case $B$ recombination. If the measured value of the ratio $F_{{\rm}H\alpha}/F_{{\rm H}\beta}$
was lower than the theoretical value, then the reddening was adopted to be zero.

The intensity of strong lines was used to separate different types of emission-line objects according to their main excitation mechanism using the standard diagnostic diagram suggested by
\citet{Baldwin1981}  (Baldwin -- Phillips -- Terlevich, BPT, classification diagram). An empirical demarcation line between the star-forming (SF) and the active galactic nucleus (AGN)
spectra in the BPT diagram (solid line in panel (b) of Fig.~\ref{figure:m-8274-12703-bpt}) suggested by \citet{Kauffmann2003} can be interpreted as the upper limit of pure star-forming  spectra.  
The spectra located left (below) of the demarcation line of \citet{Kauffmann2003} are referred to as SF-like or  H\,{\sc ii} region-like spectra (blue points
in panels (a) and (b) of Fig.~\ref{figure:m-8274-12703-bpt}). The theoretical demarcation line between the star-forming and the AGN spectra suggested by \citet{Kewley2001} (dotted line in
panel (b) of Fig.~\ref{figure:m-8274-12703-bpt}) can be interpreted as the lower limit of pure AGN spectra. The spectra located right (above) of the demarcation line of \citet{Kewley2001}
are referred to as the AGN-like spectra (black points in panels (a) and (b) of Fig.~\ref{figure:m-8274-12703-bpt}). The spectra located between the demarcation lines of
\citet{Kauffmann2003} and \citet{Kewley2001} are referred to as intermediate (INT) spectra (red points   in panels (a) and (b) of Fig.~\ref{figure:m-8274-12703-bpt}).
The long-dashed line in panel (b) of Fig.~\ref{figure:m-8274-12703-bpt} is the dividing line between Seyfert galaxies and low-ionisation nuclear emission line regions (LINERs)
 defined by \citet{CidFernandes2010}. 

We estimated the oxygen abundances in the spaxels with the H\,{\sc ii} region-like spectra using the $R$ calibration from \citet{Pilyugin2016}. The oxygen abundance map and
the radial distribution for the MaNGA galaxy M-8274-12703 are shown in panels (a) and (b) of Fig.~\ref{figure:m-8274-12703-oh-map},   respectively. We also estimated the
nitrogen-to-oxygen abundance ratios using the corresponding N/O calibration from \citet{Pilyugin2016} and then determined the nitrogen abundances log(N/H) = log(O/H) + log(N/O).
The radial distributions of the oxygen and nitrogen abundances for individual spaxels in the galaxy M-8274-12703 are shown by grey points in panels (a) and (b) of 
Fig.~\ref{figure:m-8274-12703-r-x}. In order to minimise the influence of spaxels with unreliable abundances in determining the radial abundance distribution, we determined
the radial abundance distribution using not the abundances in individual spaxels, but the median values of the abundances in bins of 0.05~dex in $R/R_{25}$ (red circles in panels
(a) and (b) of Fig.~\ref{figure:m-8274-12703-r-x}). 

We verified the validity of the flat gradient in the galaxy in the following way. Panel (c) in Fig.~\ref{figure:m-8274-12703-r-x} shows the variation in intensity of the nitrogen
emission line N$_{2}$ across the disc. We used a standard notation for the nitrogen line intensities: $N_2  = I_{\rm [N\,II] \lambda 6548+ \lambda 6584} /I_{{\rm H}\beta }$.
The intensity of the N$_{2}$ line correlates with the electron temperature in the nebula \citep[e.g.][]{Pilyugin2016}, and consequently,
with the oxygen abundance. In particular, it is assumed within the framework of the N$_2$ calibration that the oxygen abundance is a function of the intensity of the N$_{2}$ line
alone \citep{Pettini2004,Marino2013}. For these calibrations, the relation between the variations in the oxygen abundance $\Delta$(log(O/H)) and in the N$_{2}$ line $\Delta$(log(N$_2$)) is
$\Delta$(log(O/H)) $\sim$ 0.5~$\Delta$(log(N$_2$)). Then the small variation in the N$_{2}$ across the disc confirms the flat metallicity gradient in the galaxy. 
We emphasise that the value of N$_{2}$ alone does not determine the accurate abundance in the  H\,{\sc ii} region, and the N/O ratio and the excitation
parameter should be taken into account \citep{Pilyugin2016, Schaefer2020}.

\subsubsection{Radial distributions of stellar ages}

\citet{Sanchez2022} estimated the luminosity-weighed ages of the stellar populations in spaxels of the MaNGA galaxies, Age$_{\star,LW}$, by ﬁtting the spaxel spectrum with a
combination of simple stellar population spectra. The Age$_{\star,LW}$ for galaxies of our sample are taken from their
catalogue\footnote{https://data.sdss.org/sas/dr17/manga/spectro/pipe3d/v3\_1\_1/3.1.1/}.
Panel (d) in Fig.~\ref{figure:m-8274-12703-r-x} shows the stellar ages Age$_{\star,LW}$ in individual spaxels (yellow points) and the median values of the stellar ages in bins of 0.05~dex
in $R/R_{25}$ (blues circles). We specified the stellar age of a galaxy by the median value of the ages in all the spaxels within the optical radius  Age$_{\star,LW,m}$ (dashed blue
line in panel (d) in Fig.~\ref{figure:m-8274-12703-r-x}).

The spectral index D$_{n}$(4000) depends on the age of the stellar population and on the metallicity. Because the variations in metallicities within each galaxy and between galaxies considered
here are small, we can expect that the index D$_{n}$(4000) is mainly a function of the age of the stellar population. The relation between the age and index  D$_{n}$(4000) for a solar
metallicity in Figure~2 from  \citet{Kauffmann2003a} is well approximated by the expression 
\begin{eqnarray}
       \begin{array}{lll}
\log(Age_{\star,D})  & = & \,\, 7.258\, X - 1.224, \,\,\,\,\, {\rm if} \,\,X > 0.0727  \\  
                 & = &     22.550\, X - 2.336,  \,\,\,\,\, {\rm if} \,\,X < 0.0727,  \\ 
     \end{array}
\label{equation:age}
\end{eqnarray}
where X = log(D$_{n}$(4000)), and the age is in billion years.  We converted the spectral index D$_{n}$(4000) for each spaxel into age using this expression.
Panel (d) in Fig.~\ref{figure:m-8274-12703-r-x} shows the stellar ages in individual spaxels Age$_{\star,D}$ (grey points), the median values of the stellar ages in bins of 0.05~dex in
$R/R_{25}$ (red circles), and the median value of the ages in all the spaxels within the optical radius Age$_{\star,D,m}$ (dashed red line in panel (d) in Fig.~\ref{figure:m-8274-12703-r-x}).

Inspection of panel (d) in Fig.~\ref{figure:m-8274-12703-r-x} shows that the behaviour of the stellar age Age$_{\star,D}$ along radius is more or less similar to that of Age$_{\star,LW}$,
although the absolute values of Age$_{\star,D}$ and Age$_{\star,LW}$ are significantly different.  Fig.~\ref{figure:aged-agelw} shows Age$_{\star,LW,m}$ as a function Age$_{\star,D,m}$ for the
galaxies of our sample. Fig.~\ref{figure:aged-agelw} shows that the value of Age$_{\star,LW,m}$ correlates  with the value of Age$_{\star,D,m}$. This consideration suggests that the use of
either  Age$_{\star,D}$ or  Age$_{\star,LW}$ results in a qualitatively similar picture, but that the quantity characteristics can be significantly different.

\subsubsection{Sample properties}

Three independent MaNGA observations (M-8274-12703, M-8256-12703, and M-8451-12704) of the same galaxy are available. The comparison of the values of the characteristics derived from
different observations of the same galaxy 
 provides the possibility of estimating the uncertainties in characteristics obtained from the MaNGA measurements. In particular, the uncertainty in the radial oxygen abundance
gradient can be estimated. Fig.~\ref{figure:m-8274-12703-oh-nh} shows the radial distributions of the oxygen (panel a) and nitrogen (panel b) abundance distributions determined using
three observations of the same galaxy.
The close inspection of Fig.~\ref{figure:m-8274-12703-oh-nh} results in two conclusions. First, there is a general agreement between the abundance distributions obtained
from different observations of the same galaxy, 
but an appreciable difference is observed in a few bins near the centre and (or) at the periphery, near the isophotal radius of the galaxy. Therefore,
 these bins should be rejected in the determination of the radial abundance gradient. Second, the determined values of the oxygen abundance gradient are equal to   
 $-0.0264$($\pm$ 0.0105)~dex/$R_{25}$ for the M-8256-12703 measurement, +0.0026($\pm$ 0.0050)~dex/$R_{25}$ for the M-8274-12703 measurement, and  $-0.0461$($\pm$ 0.0121)~dex/$R_{25}$
 for the M-8451-12704 measurement. Thus, the uncertainty in the oxygen abundance gradient determined from MaNGA measurements can be as large as 0.05~dex/$R_{25}$. 

The values of the rotation velocity and the optical radius were also estimated using three independent observations of this galaxy. The maximum deviation of the rotation velocity
from the mean value for three measurements is $\sim$7\% or $\sim$0.029 dex.  The maximum deviation of the optical radius in arcseconds from the mean value for the three measurement is $\sim$3.5\%.
The error in the distance to the galaxy also contributes to the error of the optical radius in kiloparsec. The reported errors in distances are smaller than 10\% (see below).
We adopt an uncertainty of the rotation velocity determined from the MaNGA measurement of  $\sim$7\% or $\sim$0.029~dex and an uncertainty in the optical radius in kiloparsec
of  $\sim$13.5\% or $\sim$0.055~dex.

We find eight galaxies with nearly uniform oxygen abundances over the whole galaxy in the sense that the systematic variations in the oxygen abundances across the galaxy are smaller
than $\sim$0.05~dex, galaxies with an L-type abundance distribution. The pattern of the galaxy with this abundance distribution is shown in Fig.~\ref{figure:m-8274-12703-r-x}.  
The sample of galaxies with an LS-type abundance distribution contains 23 galaxies. An example of a galaxy with this abundance distribution is shown
in Fig.~\ref{figure:m-7495-12703-r-x}.
We selected 29 galaxies with an S-type abundance distribution, where the scatter of the binned oxygen abundance around the linear O/H -- $R_{g}$ relation is smaller than
0.01~dex. An example of a galaxy with this abundance distribution is shown in Fig.~\ref{figure:m-8131-12703-r-x}.

Fig.~\ref{figure:rc-sb-schema} shows the rotation curves normalised to the maximum value of the rotation velocity and the photometric profiles for the example galaxies
with S, LS, and L gradients presented in Fig.~\ref{figure:schema}. The inspection of Fig.~\ref{figure:rc-sb-schema} shows that there are no evident differences 
between the shapes of the rotation curves (and the photometric profiles) for the galaxies with different shapes of the radial oxygen abundance distributions.

\begin{table*}
\caption[]{\label{table:general}
    General characteristics of the MaNGA galaxies of our sample 
}
\begin{center}
\begin{tabular}{ccrcrrrrrccc} \hline \hline
Galaxy                  &
$R_{25}$                 &
$PA$                    &
$i$                     &
$d$                     &
log\,$M_{\star}$          &
$R_{25}$                 &
$V_{rot}$                &
log\,SFR                &
log\,Age$_{\star,LW,m}$    &
log\,Age$_{\star,D,m}$     &
BPT                   \\
ID                     &
arcmin                 &
degree                 &
degree                 &
Mpc                    &
$M_{\sun}$              &
Kpc                    &
km~s$^{-1}$             &
$M_{\sun}$/year         &
Gyr                    &
Gyr                    &
                      \\
\hline
(1)                    &
(2)                    &
(3)                    &
(4)                    &
(5)                    &
(6)                    &
(7)                    &
(8)                    &
(9)                    &
(10)                   &
(11)                   &
(12)              \\
\hline
  7443 12705  &  0.25  &    39.9  &  64.9  &  271.1  &  10.936  &  19.91  &  210.3  &   0.295  &  -0.123  &   0.000  &   0  \\ 
  7495 12703  &  0.38  &    15.0  &  63.0  &  130.7  &  10.604  &  14.54  &  202.6  &   0.063  &   0.013  &   0.139  &   0  \\ 
  7495 12704  &  0.44  &   351.4  &  56.4  &  127.5  &  10.872  &  16.41  &  212.8  &   0.157  &   0.161  &   0.201  &   2  \\ 
  7815 12704  &  0.25  &    75.5  &  66.1  &  178.1  &  11.072  &  12.99  &  231.5  &   0.379  &   0.167  &   0.451  &   0  \\ 
  7960 12703  &  0.31  &   321.9  &  51.3  &  101.0  &  10.445  &   9.18  &  142.2  &  -0.347  &  -0.101  &   0.149  &   0  \\ 
  7968 12703  &  0.27  &   290.4  &  62.1  &  425.4  &  11.653  &  33.10  &  366.3  &   0.605  &  -0.058  &   0.437  &   2  \\ 
  8082 12701  &  0.39  &    10.2  &  50.7  &  106.5  &  10.706  &  12.11  &  186.4  &  -0.143  &  -0.225  &   0.041  &   1  \\ 
  8131 12703  &  0.20  &   270.1  &  67.1  &  177.9  &  10.374  &  10.13  &  151.8  &  -0.195  &  -0.399  &  -0.084  &   0  \\ 
  8138 12704  &  0.47  &   165.7  &  54.4  &  128.3  &  11.363  &  17.54  &  302.2  &   0.508  &   0.143  &   0.376  &   1  \\ 
  8141 12704  &  0.22  &   158.9  &  43.2  &  199.1  &  10.630  &  12.74  &  181.5  &   0.117  &  -0.266  &  -0.010  &   1  \\ 
  8144 12702  &  0.25  &    88.5  &  68.4  &  211.5  &  10.674  &  15.33  &  179.9  &   0.225  &  -0.281  &  -0.053  &   0  \\ 
  8147 12701  &  0.26  &   297.3  &  48.3  &   98.9  &   9.779  &   7.34  &  128.6  &  -0.735  &  -0.336  &  -0.057  &   0  \\ 
  8147 12703  &  0.31  &   175.3  &  30.0  &  111.5  &   0.000  &  10.16  &  212.4  &   0.413  &  -0.282  &  -0.106  &   0  \\ 
  8257 12703  &  0.28  &   113.9  &  62.8  &  107.5  &  10.517  &   8.63  &  202.5  &   0.099  &  -0.091  &   0.227  &   0  \\ 
  8258 12703  &  0.25  &   221.5  &  69.2  &  276.4  &  11.067  &  20.10  &  212.1  &   0.410  &  -0.218  &   0.110  &   1  \\ 
  8274 12703  &  0.20  &   309.3  &  42.6  &  244.9  &  11.031  &  14.13  &  256.9  &   0.022  &   0.131  &   0.459  &   1  \\ 
  8311 12701  &  0.20  &   354.7  &  37.7  &  302.2  &  11.031  &  17.73  &  213.7  &   0.369  &  -0.015  &   0.141  &   0  \\ 
  8311 12703  &  0.34  &   142.2  &  65.1  &  139.1  &  10.790  &  13.62  &  218.7  &   0.319  &  -0.100  &   0.084  &   0  \\ 
  8315 12703  &  0.21  &   121.7  &  41.2  &  326.1  &  11.579  &  19.92  &  315.1  &   0.309  &  -0.203  &   0.202  &   2  \\ 
  8320 09102  &  0.32  &   111.6  &  49.2  &  222.9  &  11.311  &  20.75  &  271.0  &   0.286  &  -0.023  &   0.218  &   2  \\ 
  8325 12705  &  0.18  &   278.4  &  47.7  &  168.4  &   9.926  &   8.98  &  136.9  &  -0.389  &  -0.158  &  -0.027  &   0  \\ 
  8326 09102  &  0.26  &    68.7  &  27.7  &  300.3  &  11.258  &  22.42  &  313.0  &   0.218  &  -0.116  &   0.109  &   2  \\ 
  8330 12701  &  0.27  &   271.0  &  47.4  &  224.0  &  10.674  &  17.59  &  188.0  &   0.242  &  -0.361  &  -0.060  &   0  \\ 
  8332 12703  &  0.21  &   159.7  &  36.9  &  376.6  &  11.498  &  23.19  &  305.8  &   0.780  &  -0.376  &   0.100  &   1  \\ 
  8443 12705  &  0.36  &    52.1  &  63.9  &  127.8  &  10.742  &  13.54  &  215.0  &   0.129  &   0.087  &   0.337  &   1  \\ 
  8448 12703  &  0.24  &   230.2  &  65.2  &  308.4  &  11.184  &  21.83  &  295.5  &   0.307  &   0.184  &   0.263  &   2  \\ 
  8449 09102  &  0.35  &   224.5  &  47.0  &   94.0  &   9.868  &   9.50  &  121.2  &  -0.186  &  -0.462  &  -0.180  &   0  \\ 
  8453 12703  &  0.24  &   148.9  &  61.8  &  263.0  &  11.035  &  18.62  &  244.0  &   0.331  &  -0.028  &   0.228  &   0  \\ 
  8454 12702  &  0.21  &   240.4  &  39.9  &  316.9  &  11.131  &  19.20  &  278.4  &   1.097  &  -0.183  &  -0.084  &   0  \\ 
  8466 12702  &  0.36  &   318.7  &  54.4  &  122.3  &  10.532  &  12.90  &  180.4  &   0.088  &  -0.155  &  -0.053  &   0  \\ 
  8466 12704  &  0.27  &   196.8  &  42.8  &  226.5  &  10.622  &  17.62  &  154.6  &   0.255  &  -0.387  &   0.050  &   0  \\ 
  8549 12702  &  0.32  &   278.7  &  50.1  &  185.1  &  11.129  &  17.01  &  263.4  &   0.117  &  -0.076  &   0.120  &   2  \\ 
  8551 12704  &  0.33  &   321.5  &  69.8  &  127.3  &  10.727  &  12.28  &  202.0  &   0.309  &   0.078  &   0.302  &   0  \\ 
  8595 12702  &  0.24  &    53.5  &  52.0  &  196.4  &  10.462  &  13.47  &  201.5  &   0.093  &  -0.357  &  -0.127  &   0  \\ 
  8600 12702  &  0.27  &   107.0  &  57.6  &  254.4  &  10.973  &  20.23  &  228.2  &   0.477  &  -0.109  &   0.061  &   2  \\ 
  8600 12705  &  0.21  &   134.2  &  53.2  &  189.1  &  10.198  &  11.37  &  152.3  &  -0.284  &  -0.223  &  -0.007  &   0  \\ 
  8613 12702  &  0.39  &    68.3  &  67.9  &  136.8  &  10.787  &  15.35  &  195.1  &  -0.187  &   0.188  &   0.266  &   2  \\ 
  8655 09102  &  0.16  &   170.4  &  45.4  &  181.5  &  10.173  &   8.23  &  146.1  &   0.049  &  -0.155  &   0.023  &   0  \\ 
  8712 12701  &  0.25  &   303.3  &  64.5  &  208.4  &  10.651  &  15.00  &  189.2  &   0.101  &  -0.119  &   0.203  &   1  \\ 
  8718 12702  &  0.24  &    15.7  &  58.4  &  161.6  &  10.853  &  11.32  &  241.9  &   0.157  &   0.019  &   0.217  &   2  \\ 
  8725 12701  &  0.20  &   223.4  &  57.7  &  212.5  &  10.897  &  12.21  &  196.0  &  -0.116  &  -0.241  &   0.278  &   1  \\ 
  8941 12705  &  0.39  &   328.5  &  49.1  &   60.6  &   9.378  &   6.83  &   95.8  &  -1.042  &  -0.419  &  -0.272  &   0  \\ 
  8952 12703  &  0.39  &    15.6  &  63.4  &  130.7  &  10.600  &  14.89  &  205.5  &   0.033  &  -0.111  &   0.084  &   0  \\ 
  8990 12703  &  0.22  &   305.4  &  38.0  &  232.2  &  10.620  &  15.08  &  183.2  &   0.382  &  -0.218  &   0.017  &   0  \\ 
  8992 09101  &  0.23  &    25.8  &  42.8  &  103.8  &   9.687  &   6.89  &  107.2  &  -0.438  &  -0.295  &  -0.182  &   0  \\ 
  8996 12705  &  0.33  &   144.5  &  59.4  &  199.8  &  11.145  &  18.99  &  275.8  &   0.337  &   0.012  &   0.296  &   2  \\ 
  9002 12704  &  0.24  &    31.4  &  57.2  &  245.1  &  11.031  &  16.81  &  250.3  &   0.166  &  -0.025  &   0.233  &   2  \\ 
  9027 12701  &  0.38  &   260.5  &  42.6  &  136.3  &  10.996  &  14.97  &  261.7  &   0.718  &  -0.307  &  -0.074  &   1  \\ 
  9031 12701  &  0.20  &   267.8  &  37.6  &  170.0  &   9.994  &   9.77  &  149.0  &  -0.347  &  -0.181  &   0.110  &   0  \\ 
  9041 12702  &  0.34  &   344.6  &  59.5  &  137.6  &  10.876  &  13.44  &  210.6  &   0.264  &  -0.192  &   0.046  &   1  \\ 
  9041 12703  &  0.25  &   143.3  &  61.2  &  237.8  &  10.871  &  17.29  &  214.7  &   0.342  &  -0.273  &  -0.019  &   2  \\ 
  9042 09101  &  0.29  &    10.9  &  56.0  &  304.0  &  11.405  &  25.64  &  291.4  &   0.775  &   0.080  &   0.364  &   1  \\ 
  9042 12702  &  0.20  &    66.1  &  51.7  &  309.3  &  11.015  &  18.14  &  237.1  &   0.761  &  -0.112  &   0.054  &   1  \\ 
  9049 12702  &  0.22  &     9.1  &  50.9  &  378.0  &  11.332  &  23.91  &  301.0  &   0.541  &  -0.062  &   0.174  &   2  \\ 
  9485 12705  &  0.38  &    32.5  &  55.2  &  134.4  &  11.082  &  14.82  &  262.8  &   0.101  &   0.024  &   0.168  &   2  \\ 
  9493 12701  &  0.28  &    79.3  &  66.4  &   67.0  &   9.531  &   5.41  &  114.0  &  -1.017  &  -0.042  &   0.228  &   0  \\ 
  9506 12704  &  0.24  &   269.4  &  63.7  &  203.3  &  10.664  &  14.09  &  166.6  &   0.223  &  -0.352  &  -0.024  &   0  \\ 
  9508 12701  &  0.24  &   240.0  &  58.5  &  392.8  &  11.816  &  27.90  &  289.8  &   0.726  &   0.161  &   0.416  &   2  \\ 
  9868 09102  &  0.31  &   183.6  &  40.1  &  303.1  &  11.456  &  27.33  &  306.2  &   0.597  &  -0.106  &   0.020  &   1  \\ 
 10001 12705  &  0.20  &   308.5  &  48.6  &  191.6  &  10.383  &  11.38  &  188.8  &   0.243  &  -0.261  &  -0.141  &   0  \\ 
                    \hline
\end{tabular}\\
\end{center}
Notes: BPT: 0 -- H\,{\sc ii} region-like spectra at the centre, 1 -- intermediate-type spectra, 2 -- AGN-like spectra. 
\end{table*}

\begin{table*}
\caption[]{\label{table:distribution}
    Parameters of the oxygen and nitrogen abundance distributions in our sample of MaNGA galaxies. 
}
\begin{center}
\begin{tabular}{ccccccccccc} \hline \hline
Galaxy ID                  &
(O/H)$_{0}$                 &
(N/H)$_{0}$                 &
$R_{b,{\rm NH}}$                     &
(O/H)$_{R_{b,{\rm NH}}}$         &
(N/H)$_{R_{b,{\rm NH}}}$         &
$R_{b,{\rm OH}}$                     &
(O/H)$_{R_{b,{\rm OH}}}$         &
(N/H)$_{R_{b,{\rm OH}}}$         &
(O/H)$_{R_{25}}$         &
(N/H)$_{R_{25}}$         \\
\hline
  7443 12705  &  8.612  &  7.860  &  0.000  &  8.612  &  7.860  &  0.000  &  8.612  &  7.860  &  8.407  &  7.251 \\ 
  7495 12703  &  8.623  &  7.974  &  0.192  &  8.612  &  7.930  &  0.397  &  8.600  &  7.777  &  8.439  &  7.325 \\ 
  7495 12704  &  8.633  &  7.960  &  0.321  &  8.619  &  7.948  &  0.542  &  8.610  &  7.835  &  8.541  &  7.602 \\ 
  7815 12704  &  8.586  &  7.868  &  0.000  &  8.586  &  7.868  &  1.000  &  8.567  &  7.579  &  8.567  &  7.579 \\ 
  7960 12703  &  8.607  &  7.906  &  0.000  &  8.607  &  7.906  &  0.000  &  8.607  &  7.906  &  8.399  &  7.284 \\ 
  7968 12703  &  8.621  &  7.859  &  1.000  &  8.585  &  7.768  &  1.000  &  8.585  &  7.768  &  8.585  &  7.768 \\ 
  8082 12701  &  8.659  &  8.019  &  0.000  &  8.659  &  8.019  &  0.000  &  8.659  &  8.019  &  8.426  &  7.272 \\ 
  8131 12703  &  8.598  &  7.777  &  0.000  &  8.598  &  7.777  &  0.000  &  8.598  &  7.777  &  8.334  &  7.126 \\ 
  8138 12704  &  8.594  &  7.940  &  1.000  &  8.632  &  7.822  &  1.000  &  8.632  &  7.822  &  8.632  &  7.822 \\ 
  8141 12704  &  8.656  &  7.803  &  0.307  &  8.580  &  7.779  &  0.000  &  8.656  &  7.803  &  8.408  &  7.220 \\ 
  8144 12702  &  8.545  &  7.671  &  0.404  &  8.521  &  7.599  &  0.478  &  8.517  &  7.538  &  8.320  &  7.104 \\ 
  8147 12701  &  8.538  &  7.650  &  0.000  &  8.538  &  7.650  &  0.000  &  8.538  &  7.650  &  8.289  &  7.056 \\ 
  8147 12703  &  8.605  &  7.949  &  0.000  &  8.605  &  7.949  &  0.127  &  8.604  &  7.879  &  8.482  &  7.394 \\ 
  8257 12703  &  8.604  &  7.920  &  0.000  &  8.604  &  7.920  &  1.000  &  8.584  &  7.685  &  8.584  &  7.685 \\ 
  8258 12703  &  8.592  &  7.687  &  0.000  &  8.592  &  7.687  &  0.000  &  8.592  &  7.687  &  8.457  &  7.376 \\ 
  8274 12703  &  8.626  &  7.992  &  0.441  &  8.627  &  7.969  &  1.000  &  8.628  &  7.840  &  8.628  &  7.840 \\ 
  8311 12701  &  8.640  &  8.056  &  0.312  &  8.627  &  8.022  &  0.711  &  8.611  &  7.785  &  8.531  &  7.614 \\ 
  8311 12703  &  8.645  &  7.958  &  0.000  &  8.645  &  7.958  &  0.000  &  8.645  &  7.958  &  8.447  &  7.309 \\ 
  8315 12703  &  8.637  &  7.912  &  0.533  &  8.626  &  7.928  &  0.682  &  8.623  &  7.781  &  8.498  &  7.467 \\ 
  8320 09102  &  8.690  &  8.075  &  0.000  &  8.690  &  8.075  &  0.000  &  8.690  &  8.075  &  8.397  &  7.230 \\ 
  8325 12705  &  8.617  &  7.847  &  0.000  &  8.617  &  7.847  &  0.000  &  8.617  &  7.847  &  8.288  &  7.011 \\ 
  8326 09102  &  8.643  &  7.966  &  0.353  &  8.626  &  7.951  &  0.499  &  8.619  &  7.803  &  8.434  &  7.296 \\ 
  8330 12701  &  8.646  &  7.931  &  0.108  &  8.639  &  7.919  &  0.137  &  8.637  &  7.894  &  8.348  &  7.140 \\ 
  8332 12703  &  8.622  &  7.954  &  0.670  &  8.592  &  7.803  &  0.691  &  8.591  &  7.785  &  8.528  &  7.522 \\ 
  8443 12705  &  8.601  &  7.872  &  0.454  &  8.595  &  7.867  &  1.000  &  8.588  &  7.686  &  8.588  &  7.686 \\ 
  8448 12703  &  8.703  &  8.075  &  0.000  &  8.703  &  8.075  &  0.000  &  8.703  &  8.075  &  8.441  &  7.349 \\ 
  8449 09102  &  8.571  &  7.692  &  0.000  &  8.571  &  7.692  &  0.000  &  8.571  &  7.692  &  8.211  &  6.837 \\ 
  8453 12703  &  8.618  &  7.872  &  0.510  &  8.593  &  7.817  &  0.693  &  8.584  &  7.686  &  8.498  &  7.467 \\ 
  8454 12702  &  8.635  &  7.784  &  0.000  &  8.635  &  7.784  &  0.000  &  8.635  &  7.784  &  8.401  &  7.200 \\ 
  8466 12702  &  8.600  &  7.871  &  0.339  &  8.585  &  7.783  &  0.440  &  8.580  &  7.669  &  8.300  &  7.035 \\ 
  8466 12704  &  8.654  &  8.035  &  0.000  &  8.654  &  8.035  &  0.000  &  8.654  &  8.035  &  8.340  &  7.077 \\ 
  8549 12702  &  8.610  &  7.912  &  0.721  &  8.575  &  7.658  &  0.668  &  8.594  &  7.677  &  8.474  &  7.416 \\ 
  8551 12704  &  8.604  &  7.880  &  0.000  &  8.604  &  7.880  &  0.000  &  8.604  &  7.880  &  8.528  &  7.531 \\ 
  8595 12702  &  8.626  &  7.836  &  0.000  &  8.626  &  7.836  &  0.000  &  8.626  &  7.836  &  8.385  &  7.228 \\ 
  8600 12702  &  8.685  &  8.079  &  0.000  &  8.685  &  8.079  &  0.000  &  8.685  &  8.079  &  8.405  &  7.224 \\ 
  8600 12705  &  8.589  &  7.742  &  0.000  &  8.589  &  7.742  &  0.000  &  8.589  &  7.742  &  8.388  &  7.233 \\ 
  8613 12702  &  8.735  &  8.252  &  0.000  &  8.735  &  8.252  &  0.000  &  8.735  &  8.252  &  8.414  &  7.236 \\ 
  8655 09102  &  8.545  &  7.696  &  1.000  &  8.489  &  7.481  &  1.000  &  8.489  &  7.481  &  8.489  &  7.481 \\ 
  8712 12701  &  8.649  &  7.940  &  0.000  &  8.649  &  7.940  &  0.000  &  8.649  &  7.940  &  8.427  &  7.325 \\ 
  8718 12702  &  8.611  &  7.903  &  0.780  &  8.563  &  7.606  &  0.775  &  8.566  &  7.608  &  8.450  &  7.356 \\ 
  8725 12701  &  8.636  &  7.875  &  0.405  &  8.613  &  7.889  &  0.516  &  8.607  &  7.791  &  8.436  &  7.362 \\ 
  8941 12705  &  8.505  &  7.462  &  0.000  &  8.505  &  7.462  &  0.000  &  8.505  &  7.462  &  8.007  &  6.519 \\ 
  8952 12703  &  8.620  &  7.925  &  0.217  &  8.613  &  7.924  &  0.283  &  8.611  &  7.870  &  8.451  &  7.285 \\ 
  8990 12703  &  8.634  &  7.995  &  0.000  &  8.634  &  7.995  &  0.000  &  8.634  &  7.995  &  8.486  &  7.442 \\ 
  8992 09101  &  8.557  &  7.640  &  0.000  &  8.557  &  7.640  &  0.000  &  8.557  &  7.640  &  8.252  &  6.958 \\ 
  8996 12705  &  8.593  &  7.950  &  0.822  &  8.522  &  7.574  &  0.744  &  8.564  &  7.610  &  8.425  &  7.277 \\ 
  9002 12704  &  8.707  &  8.109  &  0.000  &  8.707  &  8.109  &  0.000  &  8.707  &  8.109  &  8.403  &  7.263 \\ 
  9027 12701  &  8.639  &  8.035  &  0.000  &  8.639  &  8.035  &  0.000  &  8.639  &  8.035  &  8.459  &  7.328 \\ 
  9031 12701  &  8.563  &  7.875  &  0.000  &  8.563  &  7.875  &  0.240  &  8.564  &  7.669  &  8.286  &  7.017 \\ 
  9041 12702  &  8.663  &  8.004  &  0.000  &  8.663  &  8.004  &  0.000  &  8.663  &  8.004  &  8.424  &  7.271 \\ 
  9041 12703  &  8.612  &  7.856  &  0.517  &  8.588  &  7.752  &  0.593  &  8.584  &  7.695  &  8.476  &  7.389 \\ 
  9042 09101  &  8.635  &  7.957  &  0.417  &  8.617  &  7.932  &  1.000  &  8.593  &  7.588  &  8.593  &  7.588 \\ 
  9042 12702  &  8.617  &  7.853  &  0.442  &  8.583  &  7.822  &  0.655  &  8.566  &  7.657  &  8.465  &  7.390 \\ 
  9049 12702  &  8.622  &  7.860  &  0.502  &  8.600  &  7.853  &  0.587  &  8.596  &  7.790  &  8.500  &  7.481 \\ 
  9485 12705  &  8.648  &  7.949  &  0.239  &  8.637  &  7.966  &  0.455  &  8.628  &  7.807  &  8.477  &  7.405 \\ 
  9493 12701  &  8.633  &  7.809  &  0.000  &  8.633  &  7.809  &  0.000  &  8.633  &  7.809  &  8.324  &  7.159 \\ 
  9506 12704  &  8.601  &  7.804  &  0.000  &  8.601  &  7.804  &  0.000  &  8.601  &  7.804  &  8.319  &  7.042 \\ 
  9508 12701  &  8.669  &  8.017  &  0.000  &  8.669  &  8.017  &  0.000  &  8.669  &  8.017  &  8.514  &  7.532 \\ 
  9868 09102  &  8.616  &  7.931  &  0.443  &  8.600  &  7.838  &  0.597  &  8.594  &  7.684  &  8.417  &  7.281 \\ 
 10001 12705  &  8.583  &  7.855  &  0.108  &  8.581  &  7.833  &  0.197  &  8.579  &  7.769  &  8.398  &  7.195 \\ 

 \hline
\end{tabular}\\
\end{center}
Notes. \\
Abundances are in units 12+log(X/H) \\
Radii are the fractional radii normalised to the optical radius $R_{25}$. 
\end{table*}

Table~\ref{table:general} lists the general characteristics of each galaxy. 
  The first column gives the MaNGA name for each galaxy.
  The isophotal radius $R_{25}$ in arcminutes of each galaxy is reported in Column 2. 
  The position angle and inclination are listed in Columns 3 and 4.
  The distance is reported in Column 5. The distances to the galaxies were adopted from the NASA/IPAC Extragalactic Database ({\sc ned})\footnote{The NASA/IPAC Extragalactic Database
({\sc ned}) is operated by the Jet Populsion Laboratory, California Institute of Technology, under contract with the National Aeronautics and Space Administration.
{\tt http://ned.ipac.caltech.edu/}}.  The {\sc ned} distances use flow corrections for Virgo, the Great Attractor, and Shapley Supercluster infall (adopting a
cosmological model with $H_{0} = 73$ km/s/Mpc, $\Omega_{m} = 0.27$, and $\Omega_{\Lambda} = 0.73$). The reported errors in distances are smaller than 10\%.  
The stellar mass is reported in Column 6. We chose the spectroscopic $M_{sp}$ masses of the Sloan Digital Sky Survey (SDSS) 
and BOSS \citep[BOSS stands for the Baryon Oscillation Spectroscopic Survey in SDSS-III, see][]{Dawson2013}.  The spectroscopic masses were taken from the
table {\sc stellarMassPCAWiscBC03}, and were determined using the Wisconsin method \citep{Chen2012} with the stellar population synthesis models from \citet{Bruzual2003}.
The reported errors in the values of the stellar mass are usually within 0.15 -- 0.2~dex. 
  The isophotal radius in kiloparsec, estimated from the data in Columns 2 and 5, is listed in Column 7. 
  The rotation velocity is given in Column 8.
  The star formation rate is listed in Column 9. 
  The median value of the luminosity-weighted stellar ages, Age$_{\star,LW,m}$, is reported in Column~10. The ages for individual spaxels were taken from \citet{Sanchez2022},
  and the median value was determined for all the spaxels within the optical radius.
  The median value of the stellar ages estimated using the D$_{n}$(4000) index, Age$_{\star,D,m}$, (see Section 2.2.3) is reported in Column 11.
  The BPT type of radiation at the centre of the galaxy (0 – SF, 1 – intermediate, and 2 – AGN) is given in Column 12.

Table~\ref{table:distribution} lists the parameters of the oxygen and nitrogen distributions in our sample of MaNGA galaxies. 
The first column gives the MaNGA name for each galaxy.
The central oxygen 12+log(O/H)$_{0}$ and nitrogen 12+log(O/H)$_{0}$ abundances (intersect values) are listed in Columns 2 and 3.
The fractional radius of the inner zone of the constant level of the nitrogen abundance $R_{b,{\rm NH}}$, oxygen 12+log(O/H)$_{R_{b,{\rm NH}}}$ and nitrogen 12+log(N/H)$_{R_{b,{\rm NH}}}$ abundances at
this radius are reported in Columns 4, 5, and 6. 
The fractional radius of the inner zone of the constant level of the oxygen abundance $R_{b,OH}$, oxygen 12+log(O/H)$_{R_{b,{\rm OH}}}$ and nitrogen 12+log(N/H)$_{R_{b,{\rm OH}}}$ abundances at
this radius are reported in Columns 7, 8, and 9.
The value of $R_{b,{\rm OH}}$ = 0 is for galaxies with an S-type radial abundance distribution, and the value of $R_{b,{\rm OH}}$ = 1 is for galaxies with an L-type radial abundance distribution. 
The oxygen 12+log(O/H)$_{R_{25}}$ and nitrogen 12+log(N/H)$_{R_{25}}$ abundances at the isophotal radius (intersec values)  are listed in Columns 10 and 11.

\section{Discussion}

We compare and discuss the properties of the selected galaxies with S-, LS-, and L-type abundance distributions. To consider the relations between macroscopic
characteristics of galaxies (stellar mass, rotation velocity, radius, and SFR), we used the entire sample of 136 galaxies.
Three independent MaNGA observations of one galaxy in our sample are available. Consequently, this galaxy has three MaNGA names (M-8274-12703, M-8256-12703, and M-8451-12704).
  We include this galaxy in the further analysis one time, the M-8274-12703 observation (Table~\ref{table:distribution}).    
We also compare the properties of the Milky Way and our sample of galaxies. The Milky Way characteristics were taken from \citet{Pilyugin2023}, unless otherwise stated.

\begin{figure}
\resizebox{1.00\hsize}{!}{\includegraphics[angle=000]{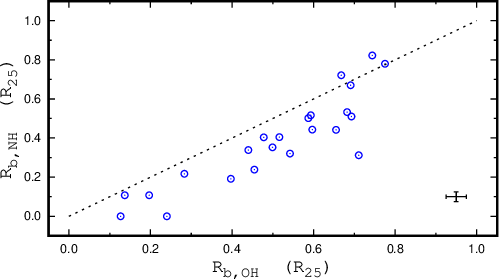}}
\caption{
  Radius of the region with a constant level of the nitrogen abundance $R_{b,{\rm NH}}$ as a function of radius  of the region with a constant level of the oxygen abundance $R_{b,{\rm OH}}$ for galaxies
  with an LS-type abundance distribution (circles). The line marks the one-to-one correspondence. The cross indicates the error bars (see text). 
}
\label{figure:rbo-rbn}
\end{figure}

\begin{figure}
\resizebox{1.00\hsize}{!}{\includegraphics[angle=000]{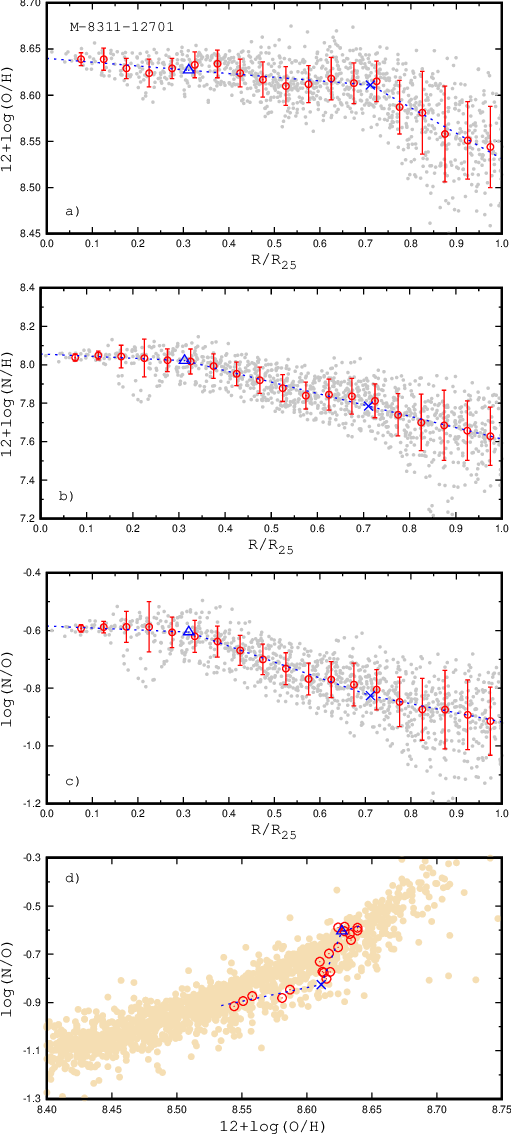}}
\caption{
  M-8311-12701 is an example of a galaxy with an LS-type abundance distribution. 
  {\em Panel} {\bf a:} Radial oxygen abundance distribution. The grey points denote the abundances for the individual spaxels, the red circles are the median values of the O/H in bins
  of 0.05~dex in the fractional radius $R/R_{25}$, and the bars show the scatter in the O/H about the median value in the bins. The line shows the adopted relation for the radial abundance
  distribution. The abundance at the radius $R_{b,{\rm OH}}$ is marked by the cross and at the radius $R_{b,{\rm NH}}$ by the triangle. 
  {\em Panel} {\bf b:} Same as panel (a), but for nitrogen abundance. 
  {\em Panel} {\bf c:} Same as panel (a), but for the N/O ratio. 
  {\em Panel} {\bf d:} N/O vs O/H diagram. The yellow points  denote H\,{\sc ii} regions in nearby galaxies (compilation in \citet{Pilyugin2016}). The other designations are the same
  as in panel (a).
}
\label{figure:m-8311-12701-oh-no}
\end{figure}

\begin{figure}
\resizebox{1.00\hsize}{!}{\includegraphics[angle=000]{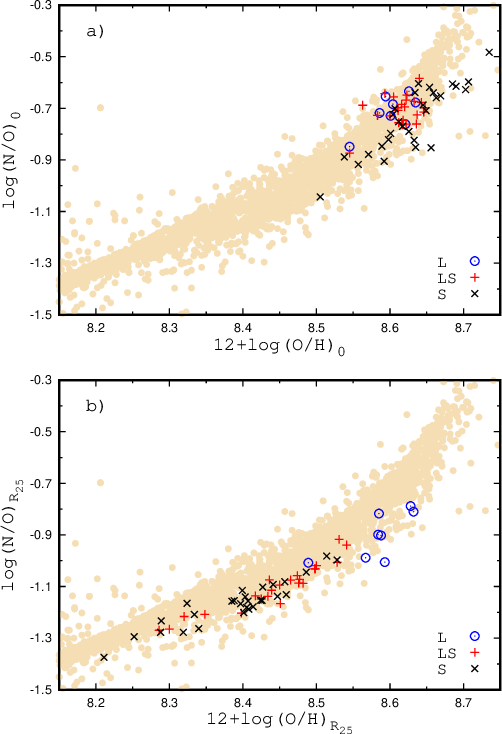}}
\caption{
  N/O vs O/H diagram for the abundances at the centres ({\em panel} {\bf a}) and at the optical radii ({\em panel} {\bf b}).
  The circles designate galaxies with an L-type abundance distribution, the plus signs mark galaxies with an LS-type abundance distribution, and crosses
  are galaxies with an S-type abundance distribution. The yellow points denote H\,{\sc ii} regions in nearby galaxies (compilation in \citet{Pilyugin2016}). 
}
\label{figure:oho-nho}
\end{figure}

\subsection{Properties of the radial abundance distributions}

The LS-type abundance distribution is the most general approximation of the abundance distributions in our sample of galaxies. The abundance distributions of the L, LS,
and S types can be described within this approximation, where $R_{b,{\rm OH}}$ = 0 for galaxies with an S-type abundance distribution,  0 $<$ $R_{b,{\rm OH}}$ $<$ 1 for galaxies with an LS-type
radial abundance distribution, and $R_{b,{\rm OH}}$ = 1 for galaxies with an L-type abundance distribution.  

A prominent feature of LS-type abundances distributions is that the nitrogen abundance distribution differs from the oxygen abundance distribution in the sense that the
size of the zone in which the nitrogen abundance is constant is usually smaller than that for the oxygen abundance (Fig.~\ref{figure:rbo-rbn} and Table~\ref{table:distribution}).
We estimated the validity of the  $R_{b,{\rm OH}}$ in each galaxy with an LS-type radial abundance distribution in the following way. We determined deviations in the
median oxygen abundance in each bin from the O/H - $R$ relations for the outer and inner regions. The deviation of the median oxygen abundance at the point
in the inner region ($R < R_{b,{\rm OH}}$) from the O/H - $R$ relation for the inner region is expected to be smaller than the deviation of the abundance at this point from the O/H - $R$
relation for the outer region, and vice versa.   
For a bin in which the median oxygen abundance does not meet this condition, the absolute value of the difference $dR$ = $R_{bin}$ -  $R_{b,{\rm OH}}$  between the radial distance
of this bin  $R_{bin}$ and the radius $R_{b,{\rm OH}}$ can be used to specify an uncertainty in the $R_{b,{\rm OH}}$ value. For example, the median oxygen abundance in the bin at the radius
of $R_{g}$ = 0.725 in the disc of galaxy M-8311-12701 does not meet the above condition (panel (a) in Fig.~\ref{figure:m-8311-12701-oh-no}).  The $dR$ value for this bin in
M-8311-12701 is equal to 0.014~$R_{25}$. We find bins like this in three more galaxies
($dR$ $\sim$ 0.03~$R_{25}$) and do not find bins like this in other galaxies (19 out of 23) with an LS-type radial abundance distribution.   
We estimated the validity of $R_{b,{\rm NH}}$ in a similar way.  We found bins in which the median nitrogen abundance did not meet the above condition in nine galaxies 
and no bins like that in other galaxies (14 out of 23) with an LS-type radial abundance distribution. This consideration suggests that the typical uncertainty in the
$R_{b,{\rm OH}}$  and  $R_{b,{\rm NH}}$ does not exceed half of the bin size.

A close examination of the changes in the oxygen and nitrogen abundances with radius shows that there are three different zones in a galaxy with an LS-type abundance distribution. 
In the outer zone $R > R_{b,{\rm OH}}$, the oxygen and nitrogen abundances both increase as the radius decreases
(panels (a) and (b) in Fig.~\ref{figure:m-8311-12701-oh-no}). The N/O ratio also increases as the radius decreases (panel (c) in Fig.~\ref{figure:m-8311-12701-oh-no}). 
The binned oxygen and nitrogen abundances in this zone are located in the lower envelope of the band outlined by the  H\,{\sc ii} regions in the nearby galaxies in the N/O--O/H diagram
(panel (d) of  Fig.~\ref{figure:m-8311-12701-oh-no}).
Within the zone from $R_{b,{\rm NH}}$ to $R_{b,{\rm OH}}$, the oxygen abundance is approximately constant, while the nitrogen abundance increases as the radius 
decreases. In the inner zone $R <  R_{b,{\rm NH}}$, the oxygen and nitrogen abundances are both at an approximately constant level. The binned oxygen and nitrogen abundances in this
zone are shifted towards an upper envelope of the band outlined by the  H\,{\sc ii} regions in nearby galaxies in the N/O--O/H diagram (panel (d) of    Fig.~\ref{figure:m-8311-12701-oh-no}). 

The N/O value for a given O/H contains important information about the heavy-element enrichment history of a region, and consequently, about its star formation history.
\citet{Edmunds1978} have noted that because oxygen and nitrogen are produced in stars of different masses, there can be a significant time delay between the release
of oxygen, which is mainly produced in massive stars, and that of nitrogen, which is produced in intermediate-mass stars, into the interstellar medium. The N/O ratio of a region then becomes
an indicator of the time,  $\tau_{\rm OH}$, that has elapsed since the last episode of star formation. The N/O -- O/H (or N/H--O/H) diagram has been subject to many investigations
\citep[][among many others]{Edmunds1978,Pilyugin1992,Pilyugin1993,Gavilan2006,Pilyugin2011,Vincenzo2016,Schaefer2020,Johnson2023}.

The observed behaviour of the oxygen and nitrogen abundances with radius in the galaxy with an LS-type abundance distribution can be explained by the time delay between nitrogen and
oxygen enrichment, together with the variation in the star formation history along the radius. The lack of systematic variation with radius of the oxygen abundance at $R < R_{b,{\rm OH}}$,
panel (a) in Fig.~\ref{figure:m-8311-12701-oh-no}, together with the increase in the N/O ratio as the radius decreases, panel (c) in Fig.~\ref{figure:m-8311-12701-oh-no},   
implies that after the region has reached a high astration level some time ago, the star formation rate in the region is reduced. This region will produce relatively small
amounts of oxygen on short timescales, while the creation of nitrogen by previous generations of stars continues, driving N/O to higher values (panels (c) and (d)
  in Fig.~\ref{figure:m-8311-12701-oh-no}.)  
The increase in the N/O ratio as the radius
decreases in the zone of constant oxygen abundance suggests that the number of the nitrogen-producing stars from previous generations, which have enough time to complete
their evolution and eject the nitrogen into the interstellar medium, increases as the radius decreases. 
This shows that the value of  $\tau_{\rm OH}(R)$ increases as the radius decreases.
The value of $\tau_{\rm OH}(R)$ becomes high enough at the radius $R_{b,{\rm NH}}$  so that a bulk of the nitrogen-producing stars from previous generations completes their evolution, the nitrogen
abundance reaches a high level at $R_{b,{\rm NH}}$ and  remains constant at $R < R_{b,{\rm NH}}$ (panels (c) and (d)   in Fig.~\ref{figure:m-8311-12701-oh-no}). 
Thus, the observed behaviour of oxygen and nitrogen abundances with radius clearly shows the
effect of the inside-out disc evolution model; the galactic centre evolves more rapidly than the regions at greater galactocentric distances \citep{Matteucci1989}.

In Fig.~\ref{figure:oho-nho} we compare the galaxy positions with different abundance distributions in the N/O--O/H diagram. Panel (a) shows the N/O--O/H
diagram for the (intersect) abundances at the centres of galaxies, and panel (b) shows a similar diagram for the (intersect) abundances at the optical radii. 
Panel (b) shows that the abundances at the optical radii of all the galaxies are located in the lower envelope of the band outlined by the  H\,{\sc ii} regions in nearby galaxies
 in the N/O--O/H diagram. This implies that the oxygen enrichment at the optical radius of each galaxy is defined by the ongoing star formation. Panel (a) of Fig.~\ref{figure:oho-nho}
shows that the abundances at the centres of most galaxies with an L- and LS-type abundance distribution are located close to the upper envelope of the band.
This shift of N/O to higher values suggests that the contribution of the star formation at the current epoch in the centres of these galaxies to the oxygen enrichment is low. In contrast, 
the abundances at the centres of many galaxies with an S-type abundance distribution are located close to the lower envelope of the band, that is, their oxygen abundances in
the centres are defined by the ongoing star formation.
 
In summary, 1) the radial nitrogen abundance distributions in galaxies with LS-type oxygen abundance distributions also show breaks at radii smaller than the O/H distribution breaks.
The observed behaviour of oxygen and nitrogen abundances with radius in these galaxies can be explained by the time delay between the nitrogen and oxygen enrichment, together with the
variation in the star formation history along the radius. These galaxies clearly show the inside-out disc evolution, that is, the galactic centre evolves more rapidly than
the regions at greater galactocentric distances.
2) The abundances at the optical radii of all the galaxies and at the centres of most galaxies with S-type abundance gradients are located in the lower envelope
of the band in the N/O -- O/H diagram. This implies that the oxygen enrichment is defined by the ongoing star formation. In contrast, the abundances at the centres of most
galaxies with L and LS gradients are shifted towards the upper envelope of the band. This suggests that these regions produce relatively small amounts of oxygen on short
timescales, while the production of nitrogen by previous generations of stars continues, driving N/O to higher values.

\subsection{Properties of the abundance distributions versus macroscopic characteristics}

\begin{figure}
\resizebox{1.00\hsize}{!}{\includegraphics[angle=000]{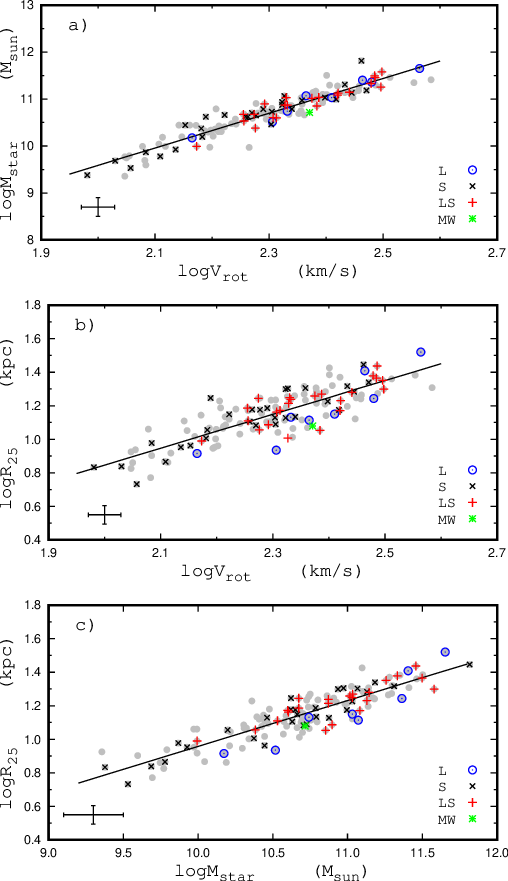}}
\caption{
  Relation between the macroscopic galaxy properties. 
  {\em Panel} {\bf a:} stellar mass as a function of rotation velocity. 
  {\em Panel} {\bf b:} isophotal radius against rotation velocity. 
  {\em Panel} {\bf c:} isophotal radius vs. stellar mass.
  The grey points in each panel designate the galaxies of the total sample, and the line is the linear best fit to these data. The cross shows the error bars. 
  The circles denote the selected galaxies with an L-type radial abundance distribution, the plus signs mark galaxies with an LS-type radial abundance distribution,
  and crosses are galaxies with an S-type radial abundance distribution. The green asterisk marks the Milky Way.
}
\label{figure:v-m-r}
\end{figure}

\begin{figure}
\centering
\resizebox{1.00\hsize}{!}{\includegraphics[angle=000]{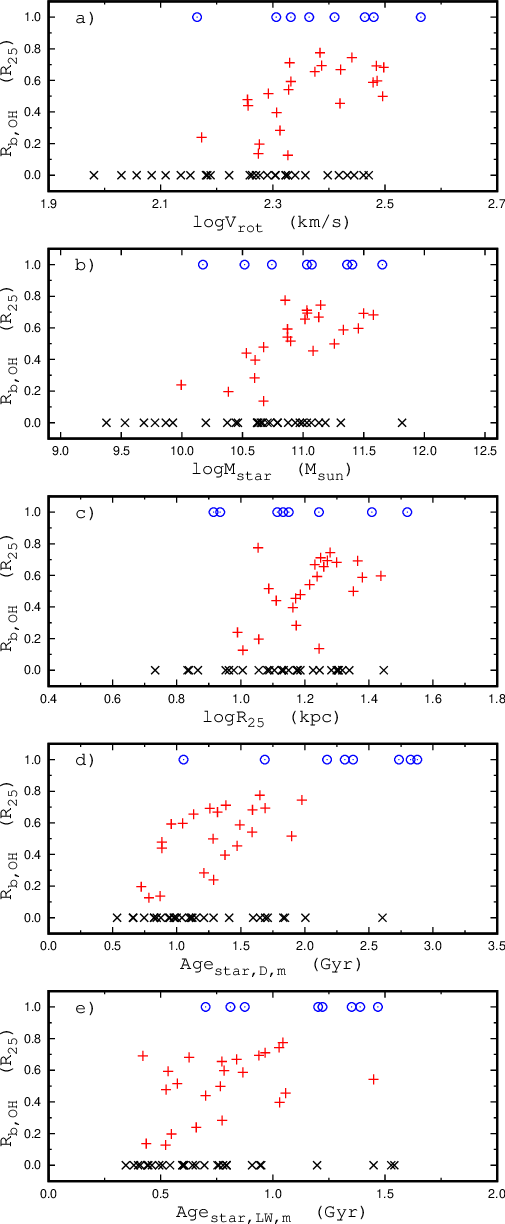}}
\caption{
  Fractional radius of the inner zone of the uniform oxygen abundance as a function of rotation velocity ({\em panel} {\bf a}), stellar mass ({\em panel} {\bf b}), optical
  radius  ({\em panel} {\bf c}),  median value of star ages estimated from the D$_{n}$(4000) index ({\em panel} {\bf d}),
  and the median value of luminosity-weighed stellar ages from \citet{Sanchez2022} ({\em panel} {\bf e}). 
  The circles denote galaxies with an L-type abundance distribution, the plus signs
  mark galaxies with an LS-type abundance distribution, and crosses are galaxies with an S-type abundance distribution. 
}
\label{figure:x-rb}
\end{figure}

\begin{figure}
\centering
\resizebox{1.00\hsize}{!}{\includegraphics[angle=000]{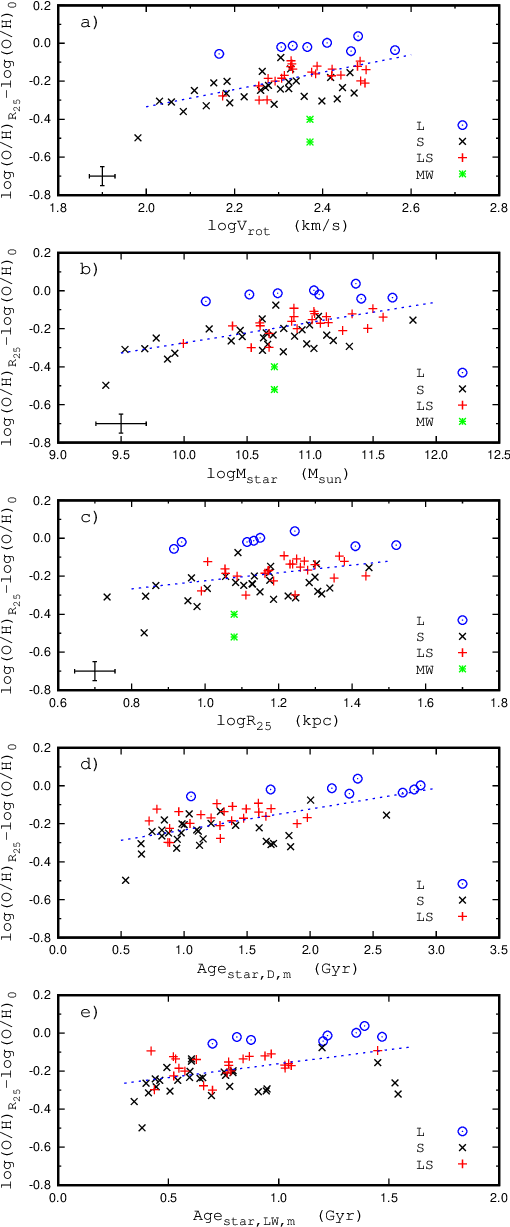}}
\caption{
  Difference between the oxygen abundances at the optical radius and at the centre (gradient in dex/$R_{25}$) as a function of rotation velocity ({\em panel} {\bf a}),
  stellar mass ({\em panel} {\bf b}), optical radius  ({\em panel} {\bf c}),  and median value of stellar ages ({\em panels} {\bf d} and {\bf e}). The blue circles denote galaxies
  with L gradients, the red plus signs mark galaxies with LS gradients, the black crosses are galaxies with S gradients, and the green asterisks mark
  two values of the gradient for the Milky Way (see text). The line is the linear fit to all the data points excluding the Milky Way. The error bars are shown in
  the lower left corner of the panel.
}
\label{figure:x-grad}
\end{figure}

First, we considered the relations between the macroscopic characteristics for galaxies with different radial abundance distribution types. Panel (a) of Fig.~\ref{figure:v-m-r}
shows the stellar mass of a galaxy as a function of its rotation velocity, the so-called Tully-Fisher relation. Panels (b) and (c) of Fig.~\ref{figure:v-m-r} show the isophotal radius
of a galaxy as a function of its rotation velocity and stellar mass, respectively. The line in each panel is a linear fit to the data points of the total sample.
Fig.~\ref{figure:v-m-r} shows that galaxies with radial abundance distribution types S, LS, and L satisfactorily follow each relation. These galaxies cannot be
distinguished on the base of their macroscopic characteristics alone.

The Milky Way is one of the galaxies with the steepest radial oxygen abundance gradient, that is, the Milky Way is the exact opposite to galaxies with an L-type radial abundance
distribution. It is of particular interest to compare the macroscopic properties of the Milky Way and the galaxies with an L-type radial abundance distribution. The Milky Way lies below the
size--rotation velocity and size--stellar mass relations (panels (b) and (c) of Fig.~\ref{figure:v-m-r}.) This is not surprising because this was revealed and discussed in a number of studies
\citep[e.g.][]{Hammer2007,Licquia2016} (but see  \citet{McGaugh2016}). Galaxies with an L-type abundance distribution also show reduced sizes on average.
Six out of eight galaxies lie below the size--rotation velocity and size--stellar mass relations, while the two most massive galaxies lie above these relations (panels (b) and (c)
of Fig.~\ref{figure:v-m-r}).  Thus, the positions of the MW in the diagrams for the macroscopic characteristics follow those of galaxies with an L-type abundance distribution, that is,
the Milky Way and galaxies with the exact opposite radial abundance distribution cannot be distinguished using the macroscopic characteristics alone.

It is interesting to note that only one galaxy (M-7968-12703) of our galaxy sample with an L-type radial abundance distribution hosts AGN at the centre. The innermost
region of the AGN-like radiation is surrounded by a ring of intermediate radiation. This is the most massive galaxy (${\rm log}(M_{\star}/M_{\sun}) = 11.653$) of our sample.
The radiation of the circumnuclear regions of four galaxies (M-8138-12704 with a stellar mass of ${\rm log}(M_{\star}/M_{\sun}) = 11.363$, M-8274-12703 with a stellar mass of
${\rm log}(M_{\star}/M_{\sun}) = 11.031$, M-8443-12705 with a stellar mass of ${\rm log}(M_{\star}/M_{\sun}) = 10.742$, and M-9042-09101 with a stellar mass of log$(M_{\star}/M_{\sun})$ = 11.405)
is of the intermediate BPT type. The circumnuclear region of the intermediate BPT type suggests that both AGN and SF contribute to the ionising radiation at the galaxy centre,
The central areas in three O/H uniform galaxies (M-7815-12704 with a stellar mass of ${\rm log}(M_{\star}/M_{\sun}) = 11.072$, M-8257-12703 with a stellar mass of
${\rm log}(M_{\star}/M_{\sun}) = 10.517$, and  M-8655-09102 with a stellar mass of ${\rm log}(M_{\star}/M_{\sun}) = 10.173$) involve spaxels with the H\,{\sc ii}-region-like radiation alone.

Fig.~\ref{figure:x-rb} shows the fractional radius of the inner zone of the uniform oxygen abundance as a function of rotation velocity, stellar mass, optical radius, and the median value of
stellar ages. As the galactic macroscopic characteristics correlate with each other, the diagrams in the different panels of Fig.~\ref{figure:x-rb} are more or less similar.
The examination of Fig.~\ref{figure:x-rb} shows that the type of the abundance distribution in a large spiral galaxy is not related to its macroscopic characteristics. A galaxy
in the stellar mass diapason $10.5 \la {\rm log}(M_{\star}/M_{\sun}) \la 11.5$ can show an S-type abundance distribution ($R_{b,{\rm OH}} = 0$), the LS type ($0 < R_{b,{\rm OH}} < 1$), and
an L type ($R_{b,{\rm OH}} = 1$; panel (b) of Fig.~\ref{figure:x-rb}). The fractional radius of the inner zone of the uniform oxygen abundance in galaxies with an LS-type
radial abundance distribution
 correlates with the stellar mass (and other macroscopic parameters) in the sense that the value of the $R_{b,{\rm OH}}$ increases with stellar mass. However, the locations of the
galaxies with S- and L-type abundance distributions disagree with this trend. They cannot be considered as the lower (upper) end of this trend.

On the other hand, galaxies with an L-type radial abundance distribution can be considered as the final stage of the galaxy evolution with S- and LS-type radial
 abundance distributions in the following sense.   
The oxygen abundances are high at the centres of galaxies for all three radial abundance distributions (panel (a) of Fig.~\ref{figure:oho-nho}), that is, the galaxy centres
of each abundance distribution are advanced in chemical evolution near the end. The oxygen abundances at the optical radii of galaxies with S- and LS-type
abundance distributions are lower than the oxygen abundances at their centres, that is, the edges of these galaxies are less advanced in chemical evolution than their centres.
At the same time,  the oxygen abundances at the optical radii of galaxies with L-type radial abundance distributions are as high as the oxygen abundances
at their centres (Fig.~\ref{figure:oho-nho}). This shows that the edges of galaxies with L-type abundance distributions are advanced in chemical evolution to
the same extent as the galaxy centres of all three radial abundance distributions.
At the end of their evolution, galaxies with S- and LS-type radial abundance distributions turn into  galaxies with L-type radial abundance distributions.

The linear fit is not an adequate approximation for the LS-type abundance distribution (Fig.~\ref{figure:schema}) and the definition of the radial gradient in these galaxies
is not beyond question. The difference between the oxygen abundances at the optical radius and at the centre $\Delta$(O/H) = ${\rm log(O/H)}_{R_{25}} - {\rm log(O/H)}_{0}$ is used
to specify the radial abundance distribution (gradient) in the galaxies with LS-type abundance distributions. It is evident that the  $\Delta$(O/H) value coincides
with the standard definition of the gradient in
dex/$R_{25}$ in galaxies with S- and L-type abundance distributions. Fig.~\ref{figure:x-grad} shows the difference between the oxygen abundances at the optical radius and at the
centre as a function of the rotation velocity, stellar mass, optical radius, and median value of star ages for our galaxy sample. The error bars are shown in the
lower left corner of the panels. The line in each panel is the linear fit to all the data points, excluding the Milky Way. 

Fig.~\ref{figure:x-grad} shows that the gradient in the galaxy correlates with the rotation velocity, stellar mass, optical radius, and median value of stellar ages for
our galaxy sample. However, the correlations are quite weak in the sense that the scatter of the points in each diagram is large.
The mean value of the deviations of $\Delta$(O/H) from the  $\Delta$(O/H) -- $V_{rot}$ relation is 0.083~dex, from
the  $\Delta$(O/H) -- $M_{\star}$ relation, it is 0.086~dex,  from the  $\Delta$(O/H) -- $R_{25}$ relation, it is 0.096 dex, from the  $\Delta$(O/H) -- $Age_{\star,D,m}$ relation, it is
0.081~dex, and from the  $\Delta$(O/H) -- $Age_{\star,LW,m}$ relation, it is 0.090~dex. 
We note that the large scatter in these diagrams cannot be attributed to the uncertainty in the  $\Delta$(O/H) values.
By the comparing the values of $\Delta$(O/H)  based on three independent observations of the same galaxy, we estimated above that the uncertainty in the   $\Delta$(O/H)
determined from the MaNGA measurements does not exceed 0.05~dex.  Fig.~\ref{figure:x-grad} shows that the gradients in galaxies with LS-type abundance
distributions are flatter on average than in galaxies with S-type abundance distributions. In agreement with the definition, the galaxies with L-type abundance distributions
show the flattest gradients.

We show two values of the gradient in the Milky Way. Spectroscopic measurements of  H\,{\sc ii} regions in the inner part ($<4$~kpc) of  the Milky Way are not available
\citep{Arellano2020,Arellano2021}. The central oxygen abundance was estimated by extrapolating the linear O/H--$R$ relation (assuming an S-type gradient) based on the available
measurements of H\,{\sc ii} regions, which results in a high central (intersect) oxygen abundance and a steep oxygen abundance gradient of about $-0.5$~dex/$R_{25}$ \citep{Pilyugin2023}.
This value of the gradient can be considered as an upper limit (maximum value) of the gradient in the Milky Way. If the gradient flattens in the central part
of the Milky Way \citep[e.g.][]{Arellano2020}, then the extrapolation of the linear relation yields an overestimated value of the central oxygen abundance and the value of
the gradient in dex/$R_{25}$ in the Milky Way. Alternatively, we can adopt the oxygen abundances measured in the H\,{\sc ii} regions, which are nearest to the centre, as the representatives of
the central oxygen abundance (assuming an LS-type gradient). The value of the gradient estimated in this way can be considered as a lower limit (minimum value) of the oxygen abundance
gradient in the Milky Way. The maximum and minimum values of the gradient in the Milky Way are shown in  Fig.~\ref{figure:x-grad}.

The radial oxygen abundance gradient in the Milky Way is steeper than in other galaxies we considered (Fig.~\ref{figure:x-grad}). The galaxies with steep
abundance gradients are small in number \citep{Sanchez2012,Sanchez2016,GarciaBenito2015,Pilyugin2023}. We note, however, that the gradients of some galaxies are much steeper than that in the Milky Way.
For example, the oxygen abundance gradient in dex/$R_{25}$ in the well-studied galaxy M~101 is steeper by a factor of about two than in the Milky Way
\citep{Kennicutt1996,Pilyugin2001,Kennicutt2003,Croxall2016}. The galaxies with L gradients and the Milky Way show the maximum opposite shifts from the gradient--$M_{\star}$
relation. At the same time, the positions of the MW in the diagrams for the macroscopic characteristics (Fig.~\ref{figure:v-m-r}) follow those of galaxies with L-type abundance distributions. 
This confirms our conclusion that the type of the abundance distribution in a large spiral galaxy is not related to its macroscopic characteristics.

In summary, 1) there are no appreciable differences between the positions of the galaxies with S-, LS-, and L-type abundance distributions in the diagrams of the different macroscopic
characteristics. 
2) The type of the abundance distribution in a large spiral galaxy is not related to its macroscopic characteristics. In particular, the positions of the MW in the diagrams of
the macroscopic characteristics show its similarity to galaxies with L-type abundance distributions, but the MW and galaxies with L gradients show the exact opposite shifts from
the gradient--$M_{\star}$ relation.
3) Our galaxy sample shows a correlation between the O/H gradient and the galactic macroscopic characteristics, but the correlations are weak in the sense
that the scatter of the points in each diagram is large.

\subsection{Evolutionary status}

\begin{figure}
\centering
\resizebox{1.00\hsize}{!}{\includegraphics[angle=000]{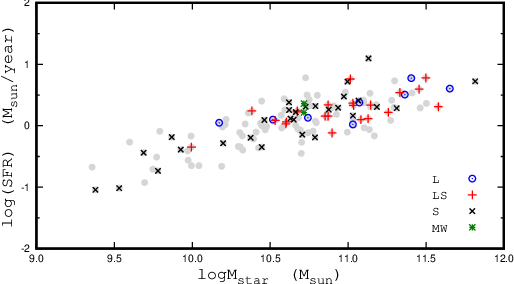}}
\caption{
  Star formation rate as a function of stellar mass. The grey points designate the galaxies of the total sample, the blue circles denote galaxies with L gradients,
  the red plus signs mark galaxies with LS gradients, and the black crosses are galaxies with S gradients. The green asterisks mark the estimates
  of the SFR in the Milky Way from  \citet{Licquia2015} and from \citet{Zari2023}  for  a Kroupa initial mass function. 
}
\label{figure:m-sfr}
\end{figure}

\begin{figure}
\centering
\resizebox{1.00\hsize}{!}{\includegraphics[angle=000]{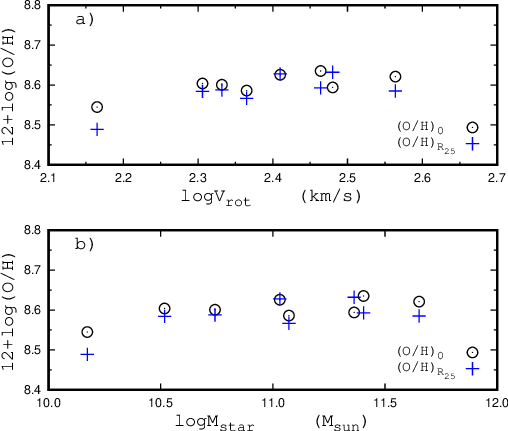}}
\caption{
  Oxygen abundance as a function of rotation velocity ({\em panel} {\bf a}) and stellar mass ({\em panel} {\bf b}) for galaxies with L gradients. The circles denote
  the oxygen abundances at the centre 12 + log(O/H)$_{0}$, and the plus signs are the oxygen abundances at the isophotal radius 12 + log(O/H)$_{R_{25}}$.  
}
\label{figure:m-oh}
\end{figure}

The diagram of the star formation rate versus stellar mass of galaxies (SFR -- $M_{\star}$) was studied in  numerous investigations 
\citep[][among many others]{Brinchmann2004,Whitaker2014,Renzini2015,Lin2017,Belfiore2018,Sanchez2018,Miranda2023,Popesso2023}.  It is well established that most galaxies lie
within two distinct bands in this diagram: the band of galaxies with a high SFR (this band is called the star-forming main sequence, or simply the main sequence) and the band of galaxies with
a low (if any) SFR (called red and dead, quiescent, or quenched sequence). The region between these two bands is populated by fewer galaxies and is usually called
the green valley. Thus, the position of a galaxy in the SFR -- $M_{\star}$ diagram indicates its evolutionary status. 

The star-forming galaxies are located within a rather narrow band in the SFR -- $M_{\star}$ diagram, with a dispersion in the SFRs of $\sim 0.2-0.3$~dex \citep[e.g.,][]{Speagle2014,Sanchez2018}.
At the same time, the differences in the value of the star formation rate in star-forming galaxies of a given stellar mass obtained in different works can be as large as a factor
of three depending on the adopted stellar mass and SFR diagnostics (e.g. the assumed stellar initial mass function and the conversion relation for estimating the star formation rate) 
\citep{Speagle2014,Miranda2023}.

We examine whether there is any difference in the positions of galaxies with different abundance distributions  in the SFR -- $M_{\star}$ diagram. We estimated the current star
formation rate from
the H$\alpha$ luminosity of a galaxy $L_{{\rm H}{\alpha}}$ using the calibration relation of \citet{Kennicutt1998} reduced by \citet{Brinchmann2004} for the Kroupa initial mass function
\citep{Kroupa2001},  
\begin{equation}
\log {\rm SFR}  = \log L_{{\rm H}{\alpha}} -41.28 . 
\label{equation:sfr}
\end{equation}
The  H$\alpha$ luminosity of a galaxy $L_{{\rm H}{\alpha}}$ was determined as the sum of the H$\alpha$ luminosities of the spaxels with  H\,{\sc ii} region-like spectra.
Fig.~\ref{figure:m-sfr} shows the SFR as a function of stellar mass for our sample of galaxies. Galaxies with different
abundance distributions are located in the same area (within the main sequence of the star-forming galaxies) in the SFR -- $M_{\star}$ diagram. This is expected because only the
galaxies where the spaxels with measured emission lines (star formation regions) are well distributed across the whole image of the galaxy were examined here because otherwise, the
abundance distributions across the galaxy cannot be determined. 

\citet{Licquia2015} determined the current star formation rate in the Milky Way based on previous measurements from the literature, assuming a Kroupa initial mass function.
They reported a value of 1.65($\pm$0.19)~$M_{\sun}$/year. 
\citet{Zari2023} mapped the stellar age distribution ($\le 1$~Gyr) across an area of 6~kpc $\times$ 6~kpc of the Galactic disc in order to constrain our Galaxy's recent star formation history. 
When this was extrapolated to the whole disc, they found an effective star formation rate over the last 10~Myr of 2.3($\pm$0.4)~$M_{\sun}$/year for the Kroupa initial mass function and 
of $\approx$ 3.3~$M_{\sun}$/year for the initial mass function that accounts for unresolved binaries.  The values of the SFR in the Milky Way obtained by  \citet{Licquia2015} and by
\citet{Zari2023} for the Kroupa initial mass function are shown by the asterisks in  Fig.~\ref{figure:m-sfr}. The examination of Fig.~\ref{figure:m-sfr} shows that the Milky Way
is located in the same area in the SFR -- $M_{\star}$ diagram as the galaxies of our sample, that is, the Milky Way and the galaxies of our sample both lie within the main sequence of
the star-forming galaxies. 

The galaxies with L-type gradients are at more advanced stages of evolution than other galaxies of our sample. Fig.~\ref{figure:m-oh} shows the oxygen abundances at the centre
$12 + {\rm log(O/H)}_{0}$ and at the isophotal radius $12 + {\rm log(O/H)}_{R_{25}}$ as a function of the rotation velocity and stellar mass for galaxies with L-type gradients.
The oxygen abundances in seven massive (log($M_{\star}/M_{\sun})\ga 10.5$)  galaxies lie in a narrow interval around $12 + {\rm log(O/H)} \sim 8.62$. We emphasise that the oxygen
abundances determined using the emission line spectra are the gas-phase abundances. \citet{Peimbert2010} estimated the dust depletion of oxygen in Galactic and extragalactic
H\,{\sc ii} regions and found that the fraction of oxygen atoms embedded in dust grains is a function of the oxygen abundance, which is about 0.12~dex for the metal-rich H\,{\sc ii}
regions. This depletion has to be considered when the total (gas + dust) oxygen abundance is derived. Then the total oxygen abundances in massive galaxies with L-type gradients are about
$12 + {\rm log(O/H)} \sim 8.74$. Empirical estimates of the oxygen yield $Y_{\rm O}$ for the metallicity scale defined by the  H\,{\sc ii} regions with $T_{e}$-based oxygen abundances result
in $Y_{\rm O} = 0.0027$ \citep{Pilyugin2004}, $Y_{\rm O} = 0.0032$ \citep{Bresolin2004}, and $Y_{\rm O} = 0.0030 \div 0.035$ \citep{Pilyugin2007}. The simple model of the galactic chemical
evolution predicts that the oxygen abundance $12 + {\rm log(O/H)}  = 8.74$ is reached at a gas mass fraction $\mu \sim 0.1$ for an oxygen yield $Y_{\rm O} = 0.0030$ and at
$\mu \sim 0.15$ for an oxygen yield $Y_{\rm O} = 0.0035$. The galaxies with L-type gradients from our sample retain sizeable reservoirs of gas to sustain the ongoing star formation
that accounts for their locations on the main sequence of star-forming galaxies. 

Thus, galaxies with different abundance distributions from our sample are located in the same area (within the main sequence of the star-forming galaxies) in the SFR -- $M_{\star}$ diagram.
The Milky Way also lies within the main sequence of the star-forming galaxies. 

\subsection{Environments}

We considered the environmental dependence of the abundance gradient in galaxy. Debris from an event of several mergers that occured between 8 and 11~Gyr ago was revealed in the Milky Way:
Gaia-Enceladus-Sausage  \citep{Belokurov2018,Helmi2018}, Thamnos \citep{Koppelman2019}, Sequoia \citep{Myeong2019}, Kraken \citep{Kruijssen2019,Kruijssen2020}, and Pontus \citep{Malhan2022}. 
The capture of the Gaia-Enceladus-Sausage is the most recent event \citep{Borre2022,Dropulic2023}.  After this, the Milky Way evolved without significant mergers for the last $\sim 10$~Gyr. 
The steep oxygen abundance gradient in the Milky Way and its evolution without significant mergers for a long time agree well with the statement in \citet{Boardman2022,Boardman2023}
that steep metallicity gradients can be found in galaxies that evolved without mergers and interactions.

We examine whether the very flat gradients in the galaxies with L-type gradients might be linked to the mergers and interactions. The merging (interacting) histories are unknown
for our sample of galaxies. We can only consider the relation between the slopes (and shapes) of the gradients in galaxies and their present-day environments. Five galaxies with
L-type gradients of our sample are in a catalogue of galaxy groups and clusters \citep{Tempel2018}. Three galaxies with L-type gradients (M-8138-12704, M-8443-12705, and M-9042-09101)
are members of galaxy groups. Two galaxies (M-8257-12703 and M-8274-12703) are classified as isolated galaxies. Nine galaxies with LS-type gradients are members of galaxy groups,
and 14 galaxies are isolated galaxies, 16 galaxies with S-type gradients are members of galaxy groups, and 11 galaxies are isolated galaxies \citep{Tempel2018}. This suggests that
the slope and shape of the radial oxygen abundance distribution in a galaxy (in particular, the nearly uniform oxygen abundances in galaxies with L-type gradients) is independent of
its present-day environment.

We note, however, that although M-8274-12703 is classified as an isolated galaxy \citep{Tempel2018}, this galaxy has a close low-mass satellite that can be seen within the field of view of
the MaNGA measurements. 
The spot in the surface brightness distribution can be seen in the image of M-8274-12703 (Fig.~\ref{figure:m-8274-12703-photometry}). The difference between the
values of the line-of-sight velocity of the centre of the galaxy and the spot is comparable to the variation in the values of the line-of-sight velocity in M-8274-12703 due
to its rotation (Fig.~\ref{figure:m-8274-12703-rc}). 

\citet{Annem2023} noted that close passages of satellites cause the formation of a substantial number of stars in the host and that the impact of the satellite perturbations
is mainly visible in the outer disc of the host. One could  expect that the advanced (chemical) evolution of the outer parts of the discs of the O/H uniform galaxies can be
attributed to perturbations by close satellites. It was noted above that the median values of the stellar ages in galaxies with L-type gradients (2-3~Gyr) are higher
than in other galaxies. This shows that the contribution of the star formation in the near past to the total star population in the discs of galaxies with L-type gradients
is low and lower than in other galaxies. The lack of significant star formation (episodes of strong star formation) in the near past can be considered as indirect
evidence in favour of the hypothesis that galaxies with L-type gradients did not experience perturbations from neighbouring galaxies during the last period of their evolution.
Therefore, if the advanced (chemical) evolution of the outer parts of the discs (i.e. the very flat oxygen abundance gradients) of galaxies with L-type gradients is caused by
perturbations from neighbouring galaxies, then these perturbations took place in the distant past.

\section{Conclusions}

We considered two sequences of well-measured spiral galaxies from the MaNGA survey with different shapes of the radial oxygen abundance distributions. The first sequence involved
galaxies whose gradient was well approximated by a single linear relation across the whole disc: the scatter of the binned oxygen abundances around the O/H--$R$ relation was smaller than
$\sim 0.01$~dex. The galaxies of this sequence are referred to as galaxies with S (slope) gradients. The second sequence contained galaxies with
LS (level-slope) gradients, where the metallicity in the inner region of the disc is at a (nearly) constant level (the gradient is flatter than $-0.05$~dex/$R_{25}$) and the gradient
is negative at larger radii. A separate subsample of galaxies with a (nearly) uniform oxygen  abundance across the whole galactic radii was selected as well.
These galaxies are called galaxies with L (level) gradients and can represent a limiting case of the two galaxy sequences we considered. Our
sample involved 29 galaxies with S-type abundance distributions, 23 galaxies with LS-type abundance distributions, and 8 galaxies
with L-type abundance distributions.  We compared properties of galaxies with different gradient shapes and that of the Milky Way. 

The radial nitrogen abundance distributions in galaxies with LS-type oxygen abundance distributions also show breaks at radii smaller than the O/H distribution breaks.
The observed behaviour of the oxygen and nitrogen abundances with radius in these galaxies can be explained by the time delay between nitrogen and oxygen enrichment, together with the
variation in the star formation history along radius. These galaxies clearly show the effect of inside-out disc evolution, which predicts that the galactic centre evolves more rapidly than
the regions at greater galactocentric distances.

We find that the abundances at the optical radii of all the galaxies and at the centres of most galaxies with S-type abundance gradients are located in the lower envelope
of the band in the N/O -- O/H diagram. This implies that the oxygen enrichment is defined by the ongoing star formation. In contrast, the abundances at the centres of most
galaxies with L and LS gradients are shifted towards an upper envelope of the band. This suggests that these regions produce relatively small amounts of oxygen on short
timescales, while the production of nitrogen by previous generations of stars continues, driving N/O to higher values. 

The abundance distribution in (large) spiral galaxies is not related to their macroscopic characteristics (rotation velocity, stellar mass, isophotal radius,
and star formation rate).  The galaxies in the stellar mass diapason $10.5 \la {\rm log}(M_{\star}/M_{\sun}) \la 11.5$ show S, LS, and L gradients. 
In particular, the Milky Way cannot be distinguished from galaxies with L gradients on the basis of the macroscopic characteristics, although the MW shows very steep abundance gradient
in contrast to extremely flat metallicity gradients in galaxies with L gradients.

The O/H gradient and the galactic macroscopic characteristics are correlated for our sample of galaxies, but the correlations are weak in the sense
 that the scatter of the points in each diagram  is large.

The galaxies with different abundance distributions in our sample and the Milky Way are located within the main sequence of the star-forming galaxies in the SFR--$M_{\star}$ diagram.

The slope and shape of the radial oxygen abundance distribution in a galaxy (in particular, the nearly uniform oxygen abundances in galaxies with L gradients) is independent of its
present-day environment. If the very flat oxygen abundance gradients in galaxies with L gradients are caused by perturbations from neighbouring galaxies, then these
perturbations had to take place in the distant past.

\section*{Acknowledgements}

We are grateful to the referee for his/her constructive comments. \\ 
L.S.P acknowledges support from the Research Council of Lithuania (LMTLT) (grant no. P-LU-PAR-23-28). \\
This research has made use of the NASA/IPAC Extragalactic Database (NED), which
is funded by the National Aeronautics and Space Administration and operated by
the California Institute of Technology.  \\
Funding for SDSS-III has been provided by the Alfred P. Sloan Foundation,
the Participating Institutions, the National Science Foundation,
and the U.S. Department of Energy Office of Science.
The SDSS-III web site is http://www.sdss3.org/. \\
Funding for the Sloan Digital Sky Survey IV has been provided by the
Alfred P. Sloan Foundation, the U.S. Department of Energy Office of Science,
and the Participating Institutions. SDSS-IV acknowledges
support and resources from the Center for High-Performance Computing at
the University of Utah. The SDSS web site is www.sdss.org. \\
SDSS-IV is managed by the Astrophysical Research Consortium for the 
Participating Institutions of the SDSS Collaboration including the 
Brazilian Participation Group, the Carnegie Institution for Science, 
Carnegie Mellon University, the Chilean Participation Group,
the French Participation Group, Harvard-Smithsonian Center for Astrophysics, 
Instituto de Astrof\'isica de Canarias, The Johns Hopkins University, 
Kavli Institute for the Physics and Mathematics of the Universe (IPMU) / 
University of Tokyo, Lawrence Berkeley National Laboratory, 
Leibniz Institut f\"ur Astrophysik Potsdam (AIP),  
Max-Planck-Institut f\"ur Astronomie (MPIA Heidelberg), 
Max-Planck-Institut f\"ur Astrophysik (MPA Garching), 
Max-Planck-Institut f\"ur Extraterrestrische Physik (MPE), 
National Astronomical Observatories of China, New Mexico State University, 
New York University, University of Notre Dame, 
Observat\'ario Nacional / MCTI, The Ohio State University, 
Pennsylvania State University, Shanghai Astronomical Observatory, 
United Kingdom Participation Group,
Universidad Nacional Aut\'onoma de M\'exico, University of Arizona, 
University of Colorado Boulder, University of Oxford, University of Portsmouth, 
University of Utah, University of Virginia, University of Washington, University of Wisconsin, 
Vanderbilt University, and Yale University.

\end{document}